\documentclass[aps,prd,superscriptaddress,preprintnumbers,reprint,nofootinbib,showpacs]{revtex4-1}
\pdfoutput=1

\usepackage{graphicx}
\usepackage{epsfig}
\usepackage{amssymb}

\usepackage{amsmath}
\usepackage{amsfonts}
\usepackage{fancyhdr}
\usepackage[utf8]{inputenc} 
\usepackage{amsmath}
\usepackage{anysize}
\usepackage{verbatim} 
\usepackage{float}
\usepackage{cancel} 
\usepackage{ dsfont }
\usepackage{framed}
\usepackage[titletoc]{appendix}
\usepackage{mathrsfs}
\usepackage{amsbsy}

\usepackage{vmargin}

\setpapersize{A4}
\setmargins{2.5cm}       
{1.5cm}                        
{17cm}                      
{23.42cm}                    
{10pt}                           
{1cm}                           
{0pt}                             
{2cm}                           

\def\lQ{\Lambda_{\rm QCD}}
\def\als{\alpha_s} 
     
\def\j{{\cal J}}

\def\m{{\cal M}}
\newcommand{\be}{\begin{equation}}
\newcommand{\ee}{\end{equation}}
\newcommand{\bea}{\begin{eqnarray}}
\newcommand{\eea}{\end{eqnarray}}
\newcommand{\nn}{\nonumber}

\begin{document}

\title{Heavy Quarkonium Hybrids: Spectrum, Decay and Mixing}
\author{Ruben Oncala}
\affiliation{Nikhef, Science Park 105, 1098 XG Amsterdam, The Netherland}
\author{Joan Soto}
\affiliation{Departament de F\'\i sica Qu\`antica i Astrof\'\i sica and Institut de Ci\`encies del Cosmos, 
Universitat de Barcelona, Mart\'\i $\;$ i Franqu\`es 1, 08028 Barcelona, Catalonia, Spain}

\date{\today}

\preprint{ICCUB-17-004, NIKHF-2017-005}

\begin{abstract}
We present a largely model independent analysis of the lighter Heavy Quarkonium Hybrids based on the strong coupling regime of Potential Non-Relativistic QCD (pNRQCD). We calculate the spectrum at leading order, including the mixing of static hybrid states. We use potentials that fulfill the required short and long distance theoretical constraints and fit well the available lattice data. We argue that the decay width  
to the lower lying Heavy Quarkonia
can be reliably estimated in some cases, and provide results for a selected set of decays. We also consider the mixing with Heavy Quarkonium states. 
We establish the form of the mixing potential at $O(1/m_Q)$, $m_Q$ being the mass of the heavy quarks, and work out its
short and long distance constraints. The weak coupling regime of pNRQCD and the effective string theory of QCD are used for that goal. We show that the mixing effects may indeed be important and produce large spin symmetry violations.
Most of the isospin zero XYZ states fit well in our spectrum, either as a Hybrid or standard Quarkonium candidates.
\end{abstract}


\maketitle

\section{Introduction}\label{sec:intr}

\begin{figure}[h]
\includegraphics[width=9cm,height=10cm]{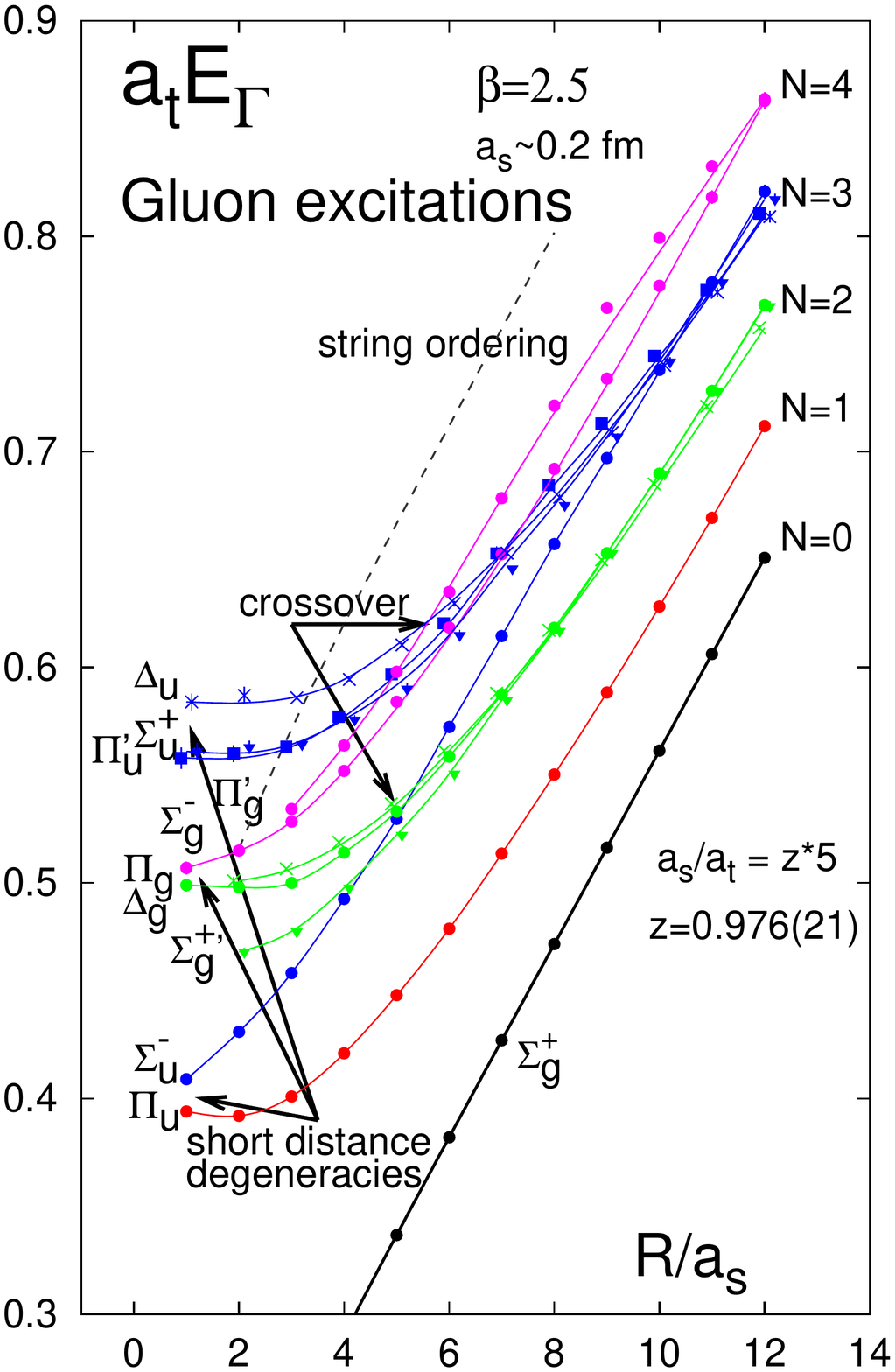}
\caption{Energy spectrum in the static limit for $n_f=0$ \cite{Juge:2002br}.}
\label{fig2}       
\end{figure}

The so called XYZ states in the charmonium and bottomonium spectrum do not fit in the usual 
potential model expectations (see \cite{Olsen:2015zcy} for a recent review). A number of models 
have been proposed to understand them, ranging from compact 
tetraquark states to just kinematical enhancements caused by the heavy-light meson pair 
thresholds. We explore here the possibility that some of these states correspond to heavy quarkonium hybrids
in a QCD based approach. Since charm and bottom masses are much 
larger than the typical QCD scale $\lQ$,  Non-Relativistic QCD (NRQCD) \cite{Caswell:1985ui,Bodwin:1994jh}
can be used for these
 states. For instance, the spectroscopy of bottomonium hybrids has been studied in lattice NRQCD in \cite{Juge:1999ie} and the production of charmonium hybrids in $B$ decays in \cite{Chiladze:1998ti}. Furthermore, if we focus on a region of the spectrum much smaller than $\lQ$, we should 
be able to build an effective theory in that region, by integrating out $\lQ$, in a way similar to the 
strong coupling regime of Potential NRQCD (pNRQCD)\cite{Brambilla:1999xf}. The static limit is relevant for such a 
construction and the spectrum in that limit is known from lattice QCD in the case of $n_f=0$ 
(no light quarks) \cite{Juge:2002br}.
 In the Born-Oppenheimer (BO) approximation,
each energy level in the static case plays the role of a potential in a Schr\"odinger equation for
 the dynamical states build on that static energy level \cite{Perantonis:1990dy}. The static spectrum is displayed
 in fig. \ref{fig2}. 

The ground state corresponds to the potential for heavy quarkonium states ($\Sigma_g^+$), namely the 
one that it is usually input in potential models. The higher levels correspond to gluonic 
excitations and are called hybrid potentials. If we are interested in states of a certain energy,
 we must in principle take into account all
the potentials below that energy, since the states build on different potentials may influence each 
other through $1/m_Q$ corrections, $m_Q$ being the mass of the heavy quarks ($Q=c,b$).
We shall focus here on the lower lying hybrid states built out of $\Pi_u$ and $\Sigma_u^-$. In addition to calculating the spectrum \cite{Juge:1999ie,Braaten:2014qka,Berwein:2015vca}, we will 
address the question on how they interact with quarkonium, namely with the states build out of $\Sigma_g^+$. 
The 
quarkonium states far below the energy of the  
hybrid states can be integrated 
out and may contribute to the decay width, whereas the
quarkonium states in the same energy range as hybrid states 
may mix with them. 
We will learn that certain hybrid states do not decay to lower lying heavy quarkonium at leading order, and that the mixing with quarkonium may induce large spin symmetry violations.
These observations will be instrumental to identify a number of XYZ states as hybrids. In fact, it turns out that most of the XYZ states can eventually be identified with either hybrids or quarkonium
in our approach. Preliminary results have been reported in \cite{Oncala:2016wlm}.

The rest of the paper is organized as follows. In Sec.~\ref{sec:hybrid} we calculate the spectrum of the lower lying hybrid states ignoring any possible mixing with other states. In Section~\ref{sec:decay} we argue that the decay width to lower lying quarkonia can be reliably estimated in some cases, and calculate it for a number of states. In Sec.~\ref{sec:mixing}, we address the mixing with quarkonium states. We establish the form of the mixing potential at ${\cal O}(1/m_Q)$, and derive the short and long distance constraints that it must fulfill using pNRQCD in the weak coupling regime \cite{Pineda:1997bj,Brambilla:1999xf} and the effective string theory of QCD respectively \cite{Luscher:2002qv,Luscher:2004ib}. We explore several interpolations for the mixing potential and recalculate the spectrum. In Section~\ref{sec:th} and Section~\ref{sec:exp} we compare our results with those of other QCD based approaches and with the experiment respectively. We also present in the latter the most likely identifications of the XYZ states as hybrids or quarkonium. Section~\ref{discussion} contains a discussion of our results. Finally, in Sec.~\ref{sec:concl} we present a short summary of the main results and conclude. Appendix \ref{quarkonium} shows our results for quarkonium. Appendix \ref{fits} provides details on how we obtain the two long distance parameters from lattice data. Appendix \ref{tensor} sets our conventions for the tensor spherical harmonics. The tables in Appendix \ref{mixspectrum} display our results for the full (quarkonium plus hybrid) charmonium and bottomonium spectrum including mixing.

\section{Spectrum}\label{sec:hybrid}

In the Born-Oppenheimer approximation, the calculation of the hybrid spectrum reduces to solving the Schr\"odinger equation
with a potential $V=V(r,\lQ)$ that has a minimum at $r= r_0\sim 1/\lQ$, $r=\vert {\bf r}\vert$, {\bf r} being the distance between the quark and the antiquark. Hence the energy of the small fluctuations about that minimum
is $E \sim \sqrt{\lQ^3/m_Q} \ll \lQ \ll m_Q$. Consequently, we are in a situation analogous to the strong coupling regime of pNRQCD in which the scale $\lQ$ is integrated out. 
It then makes sense to restrict the study to the lower lying hybrid potentials, $\Sigma_u^-$ and $\Pi_u$, since the gap to the next states is parametrically $O(\lQ)$. Specifically, from Fig. \ref{fig2} we see that the gap 
between the minimum of the $\Pi_u$ potential and the first excited potential that we neglect (${\Sigma_g^+}'$) is about $400$ MeV. Hence, for states built out of the $\Sigma_u^-$ and $\Pi_u$ potentials about $400$ MeV or more above
the lowest lying one, mixing effects with the next hybrid multiplet ( ${\Sigma_g^+}'$, $\Pi_g$, $\Delta_g$) may be relevant.

The potentials associated to $\Sigma_u^-$ and $\Pi_u$ are degenerated at short distances. In weak coupling pNRQCD \cite{Pineda:1997bj}, this is easily understood as they correspond to different projections with respect to {\bf r}
of the same operator $tr({\bf B}({\bf 0}, t) O({\bf 0},{\bf r}, t))$, where $O({\bf R},{\bf r}, t)$ is the color octet operator, ${\bf B}({\bf R}, t)$ the chromomagnetic field, and we have set the center of 
mass coordinate {\bf R}={\bf 0}. These projections have well defined transformations under the $D_{\infty h}$ group, the group of a diatomic molecule. $\hat{\bf r}{\bf B}$ corresponds to $\Sigma_u^-$
and ${\bf B}-\hat{\bf r}(\hat{\bf r}{\bf B})$ to $\Pi_u$ \cite{Brambilla:1999xf}.
It is then natural to associate to the lower lying hybrids a vectorial wave function ${\bf H}({\bf 0},{\bf r}, t)$, such that its projection to {\bf r} 
evolves with $V_{\Sigma_u^-}$ 
and its projection orthogonal to {\bf r} with $V_{\Pi_u}$. We then have the following Lagrangian density,

\bea
\label{h}
&{\cal L}={\rm tr} \left( {H^i}^\dagger\left( \delta_{ij}i\partial_0-{h_H}_{ij}\right)H_j \right)\\
&{h_{H}}_{ij}\!=\!\left(\!-\!\frac{\nabla^2}{m_Q}\!+\!V_{\Sigma_u^-}(r)\right)\delta_{ij}\!+\!\left(\delta_{ij}\!-\!\hat{r}_{i} \hat{r}_{j}\right)\left[V_{\Pi_u}(r)\!-\!V_{\Sigma_u^-}(r)\right]\,,\nonumber
\eea
where ${\bf \hat r}={\bf r}/\vert {\bf r}\vert$ and we have ignored the center of mass motion. ${\bf H}={\bf H}({\bf R},{\bf r}, t)$ is a matrix in spin space and transforms as ${\bf H} \to h_1{\bf H}h_2^\dagger$, $h_1$, $h_2 \in SU(2)$ under spin symmetry. 
${h_H}_{ij}$ above does not depend on the spin of the quarks, and hence it is invariant under spin symmetry transformations, but it does depend on the total angular momentum of the gluonic degrees of freedom ${\bf L}_g$, 
in this case $L_g=1$ as it is apparent from the vectorial character of {\bf H}. 
The symmetry properties of ${\bf H}({\bf R},{\bf r}, t)$ under parity, time reversal and charge conjugation read as follows,
\bea
P: & \, {\bf H}({\bf R},{\bf r}, t)& \rightarrow -{\bf H}(-{\bf R},-{\bf r}, t)\nn\\
T: &\, {\bf H}({\bf R},{\bf r}, t)& \rightarrow -\sigma^2{\bf H}({\bf R},{\bf r}, -t)\sigma^2\\
C: &\,  {\bf H}({\bf R},{\bf r}, t)& \rightarrow -\sigma^2{\bf H}^T({\bf R},-{\bf r}, t)\sigma^2\,,\nn
\label{symH}
\eea
where $\sigma^2$ is the second Pauli matrix. Hence the $P$ and $C$ associated to a Hybrid state with quark-antiquark orbital angular momentum $L$ and quark-antiquark spin $S$
become,
\be
P=(-1)^{L+1} \; ,\, C=(-1)^{L+S+1} \,.
\ee  

Leaving aside the spin of the quarks, it is convenient to express {\bf H} in a basis of eigenfunctions of ${\bf J}= {\bf L} + {\bf L}_g$, where {\bf L} is the orbital angular momentum of the quarks. This is achieved
using  Vector Spherical Harmonics \cite{pg},
\bea
\label{jJML}
{\bf H}({\bf r})&=&\frac{1}{r}\left(P_0^+(r)\mathcal{\bf Y}_{00}^{L=1} +\sum_{J=1}^{\infty}\sum_{M=-J}^{J}[P_J^{+}(r)\mathcal{\bf Y}_{JM}^{L=J+1}\right.\nn\\ && \left. +P_J^{0}(r)\mathcal{\bf Y}_{JM}^{L=J}+P_J^{-}(r)\mathcal{\bf Y}_{JM}^{L=|J-1|}]\right)\,.
\eea
Note that ${\bf J}$ is a conserved quantity thanks to heavy quark spin symmetry. $ \mathcal{\bf Y}_{JM}^{L}=\mathcal{\bf Y}_{JM}^{L}(\theta,\phi)$ fulfil,
\bea
&
{\bf J}^2\mathcal{\bf Y}_{JM}^{L}=J(J+1)\mathcal{\bf Y}_{JM}^{L}\, ,\quad &{\bf L}^2\mathcal{\bf Y}_{JM}^{L}=L(L+1)\mathcal{\bf Y}_{JM}^{L}\, ,\nn\\ & {\bf L}_g^2\mathcal{\bf Y}_{JM}^{L}=2\mathcal{\bf Y}_{JM}^{L}\, ,\quad & J_3\mathcal{\bf Y}_{JM}^{L}=M\mathcal{\bf Y}_{JM}^{L}\,.
\eea
The eigenvalue problem then reduces for $J\not=0$ to
\bea
\label{coph}
& 	\left[ 
 -\frac{1}{m_Q}\frac{\partial^2}{\partial r^2}\!+
\begin{pmatrix}
 		\frac{(J-1)J}{m_Qr^2}\!  & 0                    \\
 		0 & 	\frac{(J+1)(J+2)}{m_Qr^2}\!                \\
 	\end{pmatrix} +
\!V_{\Sigma_u^-}(r) +\!\right.\nn\\
 	&\left.V_q(r) \begin{pmatrix}
 		\!\frac{J+1}{2J+1}  & \frac{\sqrt{(J+1)J}}{2J+1}                    \\
 		\frac{\sqrt{(J+1)J}}{2J+1} & 	\frac{J}{2J+1}                    \\
 	\end{pmatrix}
 	\!
 	\right]\! 
	\begin{pmatrix}
 		P_J^{-}(r) \\
 		P_J^{+}(r) \\
 	\end{pmatrix}\!
	\nn
 	=\!E\!
 	\begin{pmatrix} 
 		P_J^{-}(r) \\
 		P_J^{+}(r) \\
 	\end{pmatrix} \,
	\eea

	\be
	\left(\!-\frac{1}{m_Q}\frac{\partial^2}{\partial r^2}\!+\!\frac{J(J+1)}{m_Qr^2}\!+\!V_{\Pi_u}(r)\right)P_J^{0}(r)\!=\!EP_J^{0}(r)\,,
	\label{uncoph}
	\ee
	where $V_q(r)=V_{\Pi_u}(r)-V_{\Sigma_u^-}(r)$, and for $J=0$ to
	\be
	\left(-\frac{1}{m_Q}\frac{\partial^2}{\partial r^2}+\frac{2}{m_Qr^2}+V_{\Sigma_u^-}(r)\right) P_{0}^+(r)=E P_{0}^+(r)\,.
	\label{L0}
	\ee
The equations above are equivalent to those obtained in ref. \cite{Berwein:2015vca}.	
We approximate $V_{\Pi_u}(r)$ and $V_{\Sigma_u^-}(r)$ by simple functions that have the correct behavior at short and long distances, and fit well the lattice results in fig \ref{fig2} \cite{Juge:2002br} and ref. \cite{Bali:2003jq}. For $V_{\Sigma_u^-}(r)$ it is enough to take a Cornell-like potential with the correct asymptotic behavior in order to get a good fit to data. We then take,
 \begin{equation}
\label{Vs}
V_{\Sigma_{u}^{-}}(r)=\frac{\sigma_s}{r}+\kappa_{s} r +E_s^{Q\bar{Q}}\,.
\end{equation}
The correct short and long distance behavior implies $\sigma_s=\sigma_g/8$ and $\kappa_s=\kappa_g$, where $\sigma_g$ and $\kappa_g$ are the corresponding parameters appearing in the Cornell potential for heavy quarkonium ($V_{\Sigma_{g}^{+}}(r) $), see Appendix \ref{quarkonium}. We then have,
\be
\sigma_s=0.061 \, ,\hspace{1cm} \kappa_s=0.187\,{\rm GeV}^2 \,.
\ee
The constant $E_s^{Q\bar{Q}}$ becomes then the only free parameter, which can be linked to the corresponding parameter for the heavy quarkonium
case, $E_g^{Q\bar{Q}}$ through the lattice data of ref. \cite{Juge:2002br}. Finally, $E_g^{Q\bar{Q}}$ is obtained in Appendix \ref{quarkonium}
by fitting the heavy quarkonium spectrum. We get,
\be
E_s^{c\bar{c}}=
0.559\,{\rm GeV} \,,\hspace{0.5cm} E_s^{b\bar{b}}=
0.573\, {\rm GeV} \,.
\ee
For $V_{\Pi_u}(r)$ a Cornell-like form does not fit lattice data well at intermediate distances. Hence, we take a slightly more complicated form for it,
\begin{equation}
\label{Vp}
V_{\Pi_{u}}(r) = \frac{\sigma_p}{r}\left( \frac{1+b_1r+b_2r^2}{1+a_1r+a_2r^2}\right) +\kappa_pr+E_p^{Q\bar{Q}}\,.
\end{equation}
At short distances this potential must coincide with $V_{\Sigma_{u}^{-}}(r)$ up to terms that vanish when $r\to 0$ \cite{Brambilla:1999xf}. This implies $\sigma_p=\sigma_s$ and $E_p^{Q\bar{Q}}-E_s^{Q\bar{Q}}+\sigma_p(b_1-a_1)=0$. At long distances it must be consistent with the effective string theory of QCD \cite{Luscher:2004ib},
\be
E_N(r\rightarrow\infty)=\kappa r+(\pi N-\frac{(D-2)\pi}{24})\frac{1}{r}+O(1/r^2) \,,
\ee
where $D$ is the space-time dimension and $N$ labels the energy spectrum of the string. The leading term of this formula implies
$\kappa_p=\kappa_s=\kappa$. The next-to-leading term provides the extra constraint,
\be
2\pi-\sigma_s+\frac{\sigma_pb_2}{a_2}=0 \,,
\ee
which follows from Fig. \ref{fig2} \cite{Juge:2002br}. Indeed those data show the non-trivial fact that the $V_{\Pi_{u}}(r)$ and $V_{\Sigma_{u}^{-}}(r)$ potentials at long distances correspond to the $N=1$ and $N=3$ string energy levels respectively.
Putting together all the constraints above allows to solve $a_1$, $b_1$ and $b_2$ as a function of known parameters, and $E_p^{Q\bar{Q}}$ and $a_2$, which are fitted to lattice data. We obtain,

\bea
& \sigma_p=0.061 
\,, &\kappa_p=0.187\,{\rm GeV}^2\\
 & b_1=0.06964\, {\rm GeV} 
\,, &b_2=-1.45934\, {\rm GeV}^2\nn\\
& a_1=-0.06733\, {\rm GeV}  
\,, &a_2=0.01433\, {\rm GeV}^2\nn\\
& E_p^{c\bar{c}}=
0.551\, {\rm GeV} 
\,, &E_p^{b\bar{b}}=
0.565\, {\rm GeV} \,.\nn 
\eea
The central value of lattice data and the outcome of the fits above are shown in fig. \ref{F1}, together with the potential for quarkonium $V_{\Sigma_{g}^{+}}$ discussed in the Appendix \ref{quarkonium}.

\begin{figure}[h]
	\centering 
	\includegraphics[width=0.45\textwidth]{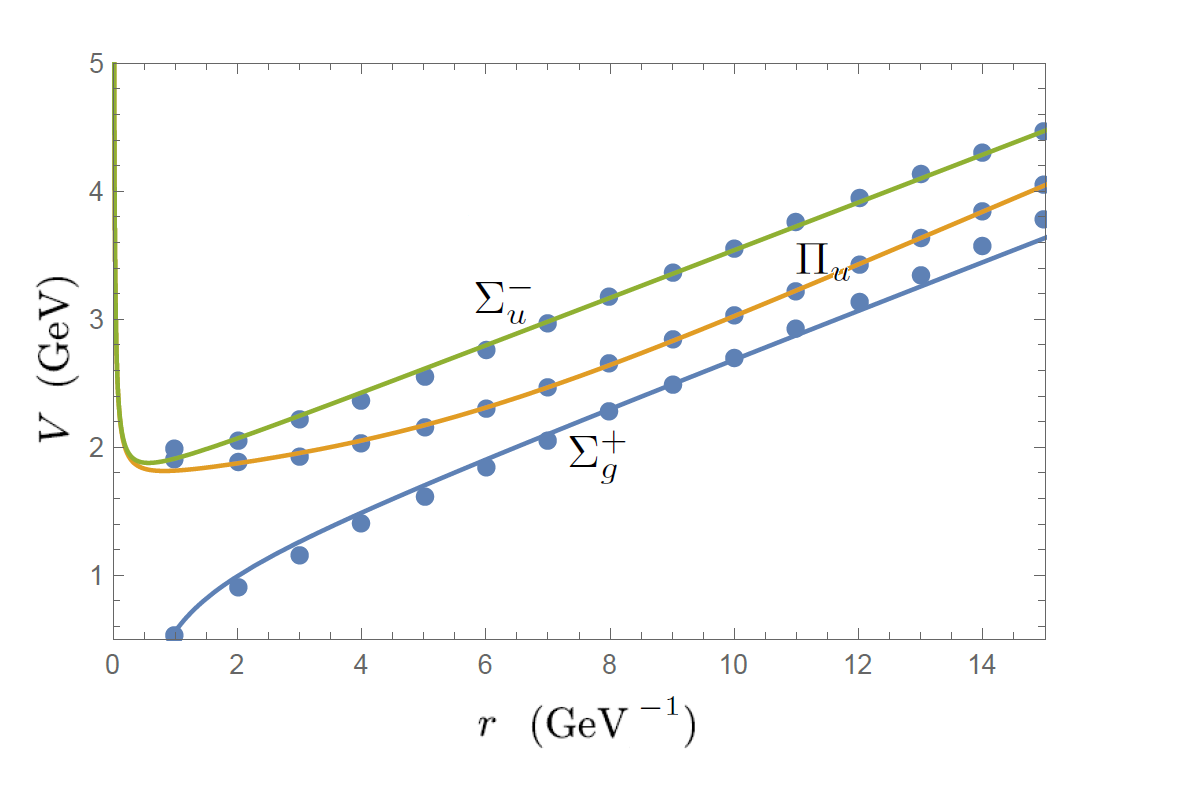}
	\caption{Our fits to the lattice results of ref. \cite{Juge:2002br} for the three lower lying B-O potentials $V_{\Sigma_{g}^{+}}, V_{\Pi_u}$  and $V_{\Sigma_u^-}$. 
	}	\label{F1}
\end{figure}  
Using the potentials above as an input we solve (\ref{coph}-\ref{L0}) and obtain the results displayed in tables \ref{cEspectrum} and \ref{bEspectrum} in terms of $M_{Q{\bar Q}g}=2m_Q+E$. Details on the code used can be found in \cite{tfm1}. We have also displayed the results in Figs. \ref{cc} and \ref{bb}, where we have included the errors discussed at the end of Sec. 
\ref{sp}.

\begin{table}[htbp]
	\centering
	\begin{tabular}{|c|c|c|c|c|c|c|}
		\hline
		&     &            &        &      $S=0$      &      $S=1$      &                             \\
		$NL_J$ & w-f & $M_{c\bar{c}}$ & $M_{c\bar{c}g}$ & $\mathcal{J}^{PC}$ & $\mathcal{J}^{PC}$ & $\Lambda^{\epsilon}_{\eta}$ \\ \hline \hline
		$1s$ & S & 3068 &  & $0^{-+}$ & $1^{--}$ & $\Sigma_g^+$ \\ 
		$2s$ & S & 3678 &  & $0^{-+}$ & $1^{--}$ & $\Sigma_g^+$ \\ 
		$3s$ & S & 4131 &  & $0^{-+}$ & $1^{--}$ & $\Sigma_g^+$ \\ 
		$1p_0$ & $P^+$ &  & 4486 & $0^{++}$ & $1^{+-}$ & $\Sigma_u^-$ \\ 
		$4s$ & S & 4512 &  & $0^{-+}$ & $1^{--}$ & $\Sigma_g^+$ \\ 
		$2p_0$ & $P^+$ &  & 4920 & $0^{++}$ & $1^{+-}$ & $\Sigma_u^-$ \\ 
		$3p_0$ & $P^+$ &  & 5299 & $0^{++}$ & $1^{+-}$ & $\Sigma_u^-$ \\ 
		$4p_0$ & $P^+$ &  & 5642 & $0^{++}$ & $1^{+-}$ & $\Sigma_u^-$ \\ \hline
		$1p$ & S & 3494 &  & $1^{+-}$ & $(0,1,2)^{++}$ & $\Sigma_g^+$ \\ 
		$2p$ & S & 3968 &  & $1^{+-}$ & $(0,1,2)^{++}$ & $\Sigma_g^+$ \\ 
		$1(s/d)_1$ & $P^{+-}$ &  & 4011 & $1^{--}$ & $(0,1,2)^{-+}$ & $\Pi_u\Sigma_u^-$ \\ 
		$1p_1$ & $P^0$ &  & 4145 & $1^{++}$ & $(0,1,2)^{+-}$ & $\Pi_u$ \\ 
		$2(s/d)_1$ & $P^{+-}$ &  & 4355 & $1^{--}$ & $(0,1,2)^{-+}$ & $\Pi_u\Sigma_u^-$ \\ 
		$3p$ & S & 4369 &  & $1^{+-}$ & $(0,1,2)^{++}$ & $\Sigma_g^+$ \\
		$2p_1$ & $P^0$ &  & 4511 & $1^{++}$ & $(0,1,2)^{+-}$ & $\Pi_u$ \\ 
		$3(s/d)_1$ & $P^{+-}$ &  & 4692 & $1^{--}$ & $(0,1,2)^{-+}$ & $\Pi_u\Sigma_u^-$ \\ 
		$4(s/d)_1$ & $P^{+-}$ &  & 4718 & $1^{--}$ & $(0,1,2)^{-+}$ & $\Pi_u\Sigma_u^-$ \\ 
		$4p$ & S & 4727 &  & $1^{+-}$ & $(0,1,2)^{++}$ & $\Sigma_g^+$ \\ 
		$3p_1$ & $P^0$ &  & 4863 & $1^{++}$ & $(0,1,2)^{+-}$ & $\Pi_u$ \\ 
		$5(s/d)_1$ & $P^{+-}$ &  & 5043 & $1^{--}$ & $(0,1,2)^{-+}$ & $\Pi_u\Sigma_u^-$ \\ 
		$5p$ & S & 5055 &  & $1^{+-}$ & $(0,1,2)^{++}$ & $\Sigma_g^+$ \\ \hline
		$1d$ & S & 3793 &  & $2^{-+}$ & $(1,2,3)^{--}$ & $\Sigma_g^+$ \\ 
		$2d$ & S & 4210 &  & $2^{-+}$ & $(1,2,3)^{--}$ & $\Sigma_g^+$ \\ 
		$1(p/f)_2$ & $P^{+-}$ &  & 4231 & $2^{++}$ & $(1,2,3)^{+-}$ & $\Pi_u\Sigma_u^-$ \\ 
		$1d_2$ & $P^0$ &  & 4334 & $2^{--}$ & $(1,2,3)^{-+}$ & $\Pi_u$ \\ 
		$2(p/f)_2$ & $P^{+-}$ &  & 4563 & $2^{++}$ & $(1,2,3)^{+-}$ & $\Pi_u\Sigma_u^-$ \\ 
		$3d$ & S & 4579 &  & $2^{-+}$ & $(1,2,3)^{--}$ & $\Sigma_g^+$ \\ 
		$2d_2$ & $P^0$ &  & 4693 & $2^{--}$ & $(1,2,3)^{-+}$ & $\Pi_u$ \\ 
		$3(p/f)_2$ & $P^{+-}$ &  & 4886 & $2^{++}$ & $(1,2,3)^{+-}$ & $\Pi_u\Sigma_u^-$ \\ 
		$4d$ & S & 4916 &  & $2^{-+}$ & $(1,2,3)^{--}$ & $\Sigma_g^+$ \\ 
		$4(p/f)_2$ & $P^{+-}$ &  & 4923 & $2^{++}$ & $(1,2,3)^{+-}$ & $\Pi_u\Sigma_u^-$ \\ 
		$3d_2$ & $P^0$ &  & 5036 & $2^{--}$ & $(1,2,3)^{-+}$ & $\Pi_u$ \\ \hline
	\end{tabular}
	\caption{ Charmonium (S) and hybrid charmonium ($P^{+-0}$)  energy spectrum computed with 
	$m_c=1.47GeV$. Masses are in MeV. States which only differ by the heavy quark spin $(S=0,1)$ are degenerated.
	$N$ is the principal quantum number, $L$ the orbital angular momentum of the heavy quarks, $J$ is $L$ plus the total angular momentum of the gluons, $S$ the spin of the heavy quarks and ${\cal J}$ is the total angular momentum. 
	For quarkonium, $J$ coincides with $L$ and it is not displayed.
	The last column shows the relevant potentials for each state. The $(s/d)_1$, $p_1$, $p_0$, $(p/f)_2$ and $d_2$ states are named $H_1$, $H_2$, $H_3$, $H_4$ and $H_5$ respectively in \cite{Berwein:2015vca}. 
	}
	
	\label{cEspectrum}
\end{table}

	\begin{table}[htbp]
		\centering
		\begin{tabular}{|c|c|c|c|c|c|c|}
			\hline 
			&     &            &        &  $S=0$ & $S=1$ & \\
			$NL_J$ & w-f & $M_{b\bar{b}}$ & $M_{b\bar{b}g}$ & $\mathcal{J}^{PC}$ & $\mathcal{J}^{PC}$ & $\Lambda^{\epsilon}_{\eta}$ \\ \hline \hline
			$1s$ & S & 9442 &  & $0^{-+}$ & $1^{--}$ & $\Sigma_g^+$ \\ 
			$2s$ & S & 10009 &  & $0^{-+}$ & $1^{--}$ & $\Sigma_g^+$ \\ 
			$3s$ & S & 10356 &  & $0^{-+}$ & $1^{--}$ & $\Sigma_g^+$ \\ 
			$4s$ & S & 10638 &  & $0^{-+}$ & $1^{--}$ & $\Sigma_g^+$ \\ 
			$1p_0$ & $P^+$ &  & 11011 & $0^{++}$ & $1^{+-}$ & $\Sigma_u^-$ \\ 
			$2p_0$ & $P^+$ &  & 11299 & $0^{++}$ & $1^{+-}$ & $\Sigma_u^-$ \\ 
			$3p_0$ & $P^+$ &  & 11551 & $0^{++}$ & $1^{+-}$ & $\Sigma_u^-$ \\ 
			$4p_0$ & $P^+$ &  & 11779 & $0^{++}$ & $1^{+-}$ & $\Sigma_u^-$ \\ \hline
			$1p$ & S & 9908 &  & $1^{+-}$ & $(0,1,2)^{++}$ & $\Sigma_g^+$ \\ 
			$2p$ & S & 10265 &  & $1^{+-}$ & $(0,1,2)^{++}$ & $\Sigma_g^+$ \\ 
			$3p$ & S & 10553 &  & $1^{+-}$ & $(0,1,2)^{++}$ & $\Sigma_g^+$ \\ 
			$1(s/d)_1$ & $P^{+-}$ &  & 10690 & $1^{--}$ & $(0,1,2)^{-+}$ & $\Pi_u\Sigma_u^-$ \\ 
			$1p_1$ & $P^0$ &  & 10761 & $1^{++}$ & $(0,1,2)^{+-}$ & $\Pi_u$ \\ 
			$4p$ & S & 10806 &  & $1^{+-}$ & $(0,1,2)^{++}$ & $\Sigma_g^+$ \\ 
			$2(s/d)_1$ & $P^{+-}$ &  & 10885 & $1^{--}$ & $(0,1,2)^{-+}$ & $\Pi_u\Sigma_u^-$ \\ 
			$2p_1$ & $P^0$ &  & 10970 & $1^{++}$ & $(0,1,2)^{+-}$ & $\Pi_u$ \\ 
			$5p$ & S & 11035 &  & $1^{+-}$ & $(0,1,2)^{++}$ & $\Sigma_g^+$ \\ 
			$3(s/d)_1$ & $P^{+-}$ &  & 11084 & $1^{--}$ & $(0,1,2)^{-+}$ & $\Pi_u\Sigma_u^-$ \\ 
			$4(s/d)_1$ & $P^{+-}$ &  & 11156 & $1^{--}$ & $(0,1,2)^{-+}$ & $\Pi_u\Sigma_u^-$ \\ 
			$3p_1$ & $P^0$ &  & 11175 & $1^{++}$ & $(0,1,2)^{+-}$ & $\Pi_u$ \\
			$6p$ & S & 11247 &  & $1^{+-}$ & $(0,1,2)^{++}$ & $\Sigma_g^+$ \\ 
			$5(s/d)_1$ & $P^{+-}$ &  & 11284 & $1^{--}$ & $(0,1,2)^{-+}$ & $\Pi_u\Sigma_u^-$ \\ \hline
			$1d$ & S & 10155 &  & $2^{-+}$ & $(1,2,3)^{--}$ & $\Sigma_g^+$ \\ 
			$2d$ & S & 10454 &  & $2^{-+}$ & $(1,2,3)^{--}$ & $\Sigma_g^+$ \\ 
			$3d$ & S & 10712 &  & $2^{-+}$ & $(1,2,3)^{--}$ & $\Sigma_g^+$ \\
			$1(p/f)_2$ & $P^{+-}$ &  & 10819 & $2^{++}$ & $(1,2,3)^{+-}$ & $\Pi_u\Sigma_u^-$ \\ 
			$1d_2$ & $P^0$ &  & 10870 & $2^{--}$ & $(1,2,3)^{-+}$ & $\Pi_u$ \\ 
			$4d$ & S & 10947 &  & $2^{-+}$ & $(1,2,3)^{--}$ & $\Sigma_g^+$ \\ 
			$2(p/f)_2$ & $P^{+-}$ &  & 11005 & $2^{++}$ & $(1,2,3)^{+-}$ & $\Pi_u\Sigma_u^-$ \\ 
			$2d_2$ & $P^0$ &  & 11074 & $2^{--}$ & $(1,2,3)^{-+}$ & $\Pi_u$ \\ 
			$5d$ & S & 11163 &  & $2^{-+}$ & $(1,2,3)^{--}$ & $\Sigma_g^+$ \\ 
			$3(p/f)_2$ & $P^{+-}$ &  & 11197 & $2^{++}$ & $(1,2,3)^{+-}$ & $\Pi_u\Sigma_u^-$ \\ 
			$3d_2$ & $P^0$ &  & 11275 & $2^{--}$ & $(1,2,3)^{-+}$ & $\Pi_u$ \\ 
			$4(p/f)_2$ & $P^{+-}$ &  & 11291 & $2^{++}$ & $(1,2,3)^{+-}$ & $\Pi_u\Sigma_u^-$ \\ \hline
		\end{tabular}
		\caption{ Bottomonium (S) and hybrid bottomonium ($P^{+-0}$)  energy spectrum computed with $m_b=4.88GeV$. 
		Masses are in MeV. States which only differ by the heavy quark spin $(S=0,1)$ are degenerated.
	$N$ is the principal quantum number, $L$ the orbital angular momentum of the heavy quarks, $J$ is $L$ plus the total angular momentum of the gluons, $S$ the spin of the heavy quarks and ${\cal J}$ is the total angular momentum. 
	For quarkonium, $J$ coincides with $L$ and it is not displayed.
	The last column shows the relevant potentials for each state. The $(s/d)_1$, $p_1$, $p_0$, $(p/f)_2$ and $d_2$ states are named $H_1$, $H_2$, $H_3$, $H_4$ and $H_5$ respectively in \cite{Berwein:2015vca}. 
	}
		
		\label{bEspectrum}
	\end{table}
\begin{widetext}
	
	\begin{figure}
\includegraphics[scale=0.85]{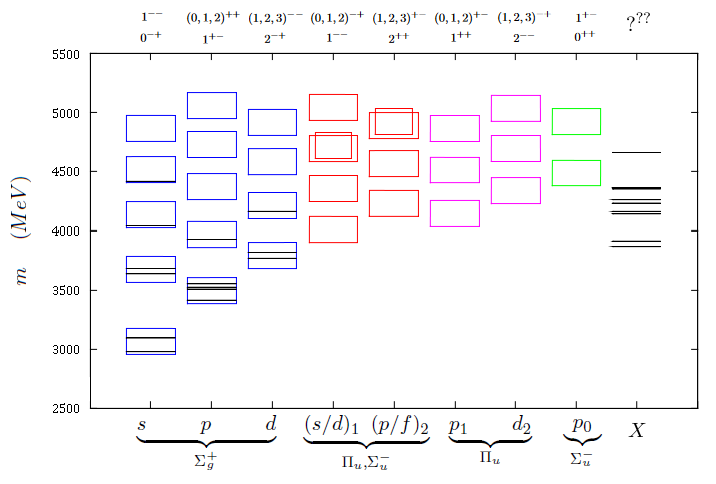}
\caption{Charmonium spectrum in Table \ref{cEspectrum}. The height of the boxes corresponds to the error estimated at the end of Sec. \ref{sp}. The states identified as quarkonium in the PDG \cite{Olive:2016xmw} are displayed in the corresponding column, whereas the states labeled as $X$ in the PDG \cite{Olive:2016xmw}  are displayed in a separated column. The box assignment of the latter is discussed in Sec. \ref{sec:exp}.   }
\label{cc}       
\end{figure}

\begin{figure}
\includegraphics[scale=0.85]{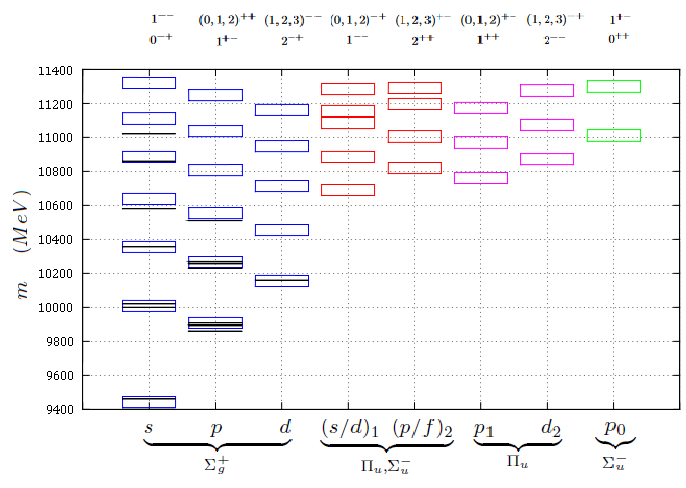}
\caption{Bottomonium spectrum in Table \ref{bEspectrum}. The height of the boxes corresponds to the error estimated at the end of Sec. \ref{sp}. The states identified as quarkonium in the PDG \cite{Olive:2016xmw} are displayed in the corresponding column.  }
\label{bb}       
\end{figure}

\end{widetext}

\section{Decay}\label{sec:decay}

\begin{figure}[H]
		\centering 
		\includegraphics[width=0.45\textwidth]{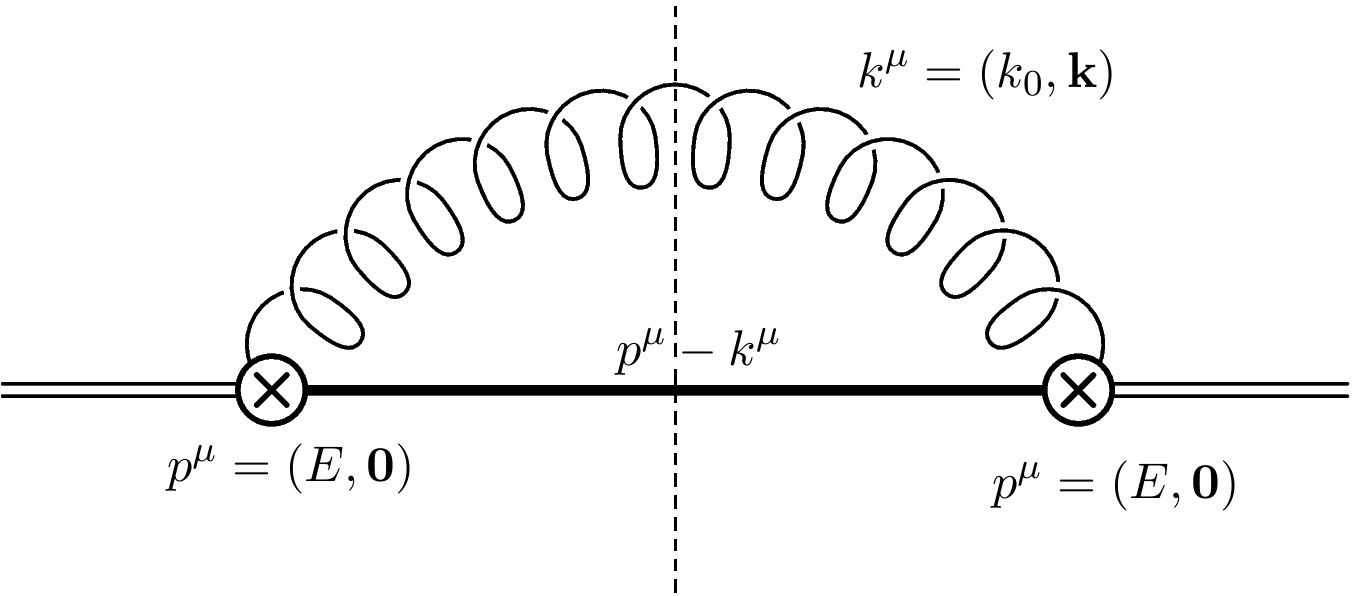}
		\caption{ The octet field self-energy diagram in weak coupling pNRQCD \cite{Brambilla:2004jw}. Double line represents the octet propagator, while single lines represent the singlet propagator. The curly line stands for the gluon propagator and the crossed circles for chromoelectric dipole vertices. The expectator gluons that make up the physical state together with the octet field are not displayed.}
		\label{decay}	
	\end{figure}

Since we are interested in the lower lying hybrid states, it is enough for us to consider an effective theory  for energy fluctuations much smaller than $\lQ$ around those states. 
The energy gap to the lower lying quarkonium states is greater than $\lQ$. Hence those states can be integrated out, which will give rise to an imaginary potential $\Delta V$, 
which in turn will produce the semi-inclusive decay width for a hybrid state to decay into any quarkonium state, $\Gamma_{H_m\to S}=-2$ $\langle H_m|\text{Im}\Delta V|H_m\rangle$. 
This is much in the same way as integrating out hard gluons in QCD produces operators with imaginary matching coefficients in NRQCD \cite{Bodwin:1994jh}, which give rise to the total decay width of a given
 quarkonium state to light degrees of freedom. Furthermore, if we assume that the energy gap to a given quarkonium state $S_n$, $\Delta E_{mn}$, fulfils  $\Delta E_{mn}\gg \lQ$, and that the process is 
short distance dominated, the integration for that state can be done using the weak coupling regime of pNRQCD \cite{Pineda:1997bj,Brambilla:1999xf},

	\bea
	\label{pNRQCD'}
	&&{\mathcal{L}}_{pNRQCD} = 
	{\rm Tr} \,\Big\{ {\rm S}^\dagger \left( i\partial_0 - h_s  \right) {\rm S} 
	+ {\rm O}^\dagger \left( iD_0 - h_o   \right) {\rm O} \Big\}  \nn\\ 
	&& + {\rm Tr} \left\{\!   
	{\rm O}^\dagger {\bf r} \cdot g{\bf E}\,{\rm S} 
	+ \hbox{H.c.} + 
	{{\rm O}^\dagger {\bf r} \cdot g{\bf E} \, {\rm O} \over 2} +{{\rm O}^\dagger {\rm O} {\bf r} \cdot g{\bf E} \over 2} \!\right\} \nn\\
	&& \quad\quad + \cdots \, .
	\eea
	The singlet field S encodes the quarkonium states whereas the octet field O encodes the heavy quark content of the hybrid states, $h_s$ and $h_o$ are Hamiltonians containing the respective Coulomb-type potentials 
and ${\bf E}={\bf E}({\bf R},t)$ is the chromoelectric field (see \cite{Brambilla:2004jw} for details).
The leading contribution
 corresponds to the diagram in fig. \ref{decay}. We obtain,
\begin{equation}
	{\rm Im}\Delta V=-
	\frac{2}{3}\frac{\als T_F}{N_c} \sum_n r^i|S_n\rangle\langle S_n|r^i\, 
	(i\partial_t-E_n)^3\,,
	\label{imdeltaV}
	\end{equation} 
	$T_F=1/2$, $N_c=3$, and $\als=g^2/4\pi$ is the QCD strong coupling constant. $E_n$ is the energy  
of the n-th quarkonium state, $S_n$.

From the expression above, we identify
\begin{equation}
	\Gamma(H_m\!\to \!S_n)\!=\!
	\frac{4}{3}\!\frac{\als T_F }{ N_c}\! \left\langle H_m|r^i|S_n\right\rangle \!\left\langle S_n|r^i|H_m\right\rangle \!\Delta E_{mn}^{\,3}\, .
	\label{dw}
	\end{equation}
$\Delta E_{mn}=E_m-E_n$, $E_m$ being the energy of the hybrid state.
At this order, the decays respect heavy quark spin symmetry, and hence the spin of the heavy quarks must be the same in the initial  hybrid state and in the final quarkonium state. In addition, 	
a selection rule derived from this formula is that hybrid states with $L=J$ do not decay to lower lying quarkonium. This selection rule will be instrumental later on to select hybrid candidates among competing XYZ states. For the allowed decays, the numerical values of the decay widths are given in table \ref{decayW}. We have only displayed numbers that can be reliably estimated, namely that $\Delta E_{mn}$ is large enough and that $\langle H_m|r^i|S_n\rangle$ is small enough so that the weak coupling regime of pNRQCD is sensible, see the table caption for details. We have taken the energies and wave functions for quarkonium  and for hybrids from Appendix \ref{quarkonium} and from the previous section respectively. 
The errors account for the fact that the quarkonium spectrum in (\ref{imdeltaV}) is meant to be calculated in the weak coupling regime (Coulomb type bound states) whereas we actually use in (\ref{dw}) the one in the strong coupling regime.

\begin{table}[htbp]
	\centering
	\begin{tabular}{|c|c|c|c|c|c|}
		\hline

			$NL_J \rightarrow N'L'$ & $\Delta E$ 
& $\langle r\rangle 
_{mn}$ 
 & ${\vert\Delta E \langle r\rangle 
_{mn}\vert}$ & $\als (\Delta E)$ & {$\Gamma$ (MeV) }  \\ \hline
					
		$1p_0\rightarrow 2s$ & {808} & {0.40} & {0.32} & 0.41 & {7.5(7.4)} \\ 
		$2(s/d)_1\rightarrow 1p$ & 861 & 0.63 & 0.54 & 0.39 & 22(19) \\ 
                $4(s/d)_1\rightarrow 1p$ & 1224 & 0.42 & 0.51 & 0.33 & 23(15) \\ 
		\hline
		\hline
		$1p_0\rightarrow 1s$ & {1569} & {-0.42} & {0.65} & 0.29 & {44(23)}   \\ 
		$1p_0\rightarrow 2s$ & {1002} & {0.43} & {0.43} & 0.36 & {15(9)}   \\  
		$2p_0\rightarrow 2s$ & {1290} & {-0.14} & {0.18} & 0,32 &{2.9(1.3)}   \\ 
		$2p_0\rightarrow 3s$ & {943} & {0.46} & {0.44} & 0.37 & {15(12)}   \\ 
                $4p_0\rightarrow 1s$ & {2337} & {0.27} & {0.63} & 0.25 & {53(25)}   \\
                 $4p_0\rightarrow 2s$ & {1770} & {0.23} & {0.40} & 0.28 & {18(7)}   \\
                $4p_0\rightarrow 3s$ & {1423} & {0.19} & {0.28} & 0.31 & {7.4(4.1)}   \\
		{$2(s/d)_1\rightarrow 1p$} & {977} & {0.47} & {0.46} & 0.37 & {17(8)}  \\ 
                 {$3(s/d)_1\rightarrow 1p$} & {1176} & {0.49} & {0.58} & 0.33 & {29(14)}  \\ 
                 {$3(s/d)_1\rightarrow 2p$} & {818} & {0.32} & {0.26} & 0.40 & {5(3)}  \\ 
                 {$4(s/d)_1\rightarrow 2p$} & {891} & {-0.74} & {0.66} & 0.39 & {33(25)}  \\
                 {$5(s/d)_1\rightarrow 1p$} & {1376} & {-0.31} & {0.43} & 0.31 & {18(7)}  \\ 
                 {$5(s/d)_1\rightarrow 2p$} & {1018} & {-0.41} & {0.42} & 0.36 & {14(8)}  \\
		\hline
	\end{tabular}
	\caption{Decay widths for hybrid charmonium (above) and bottomonium (below) to lower lying charmonia and bottomonia respectively. $m=NL_J$, $n=N'L'$, $\Delta E\equiv \Delta E_{mn}$ and $\Gamma$ are in MeV, and 
$\langle r\rangle _
{mn}$ in GeV$^{-1}$. $\als (\Delta E)$ is the one-loop running coupling constant at the scale $\Delta E$. We only display results
	for which $\Delta E > 800$MeV and ${\vert\Delta E \langle r\rangle _{mn}\vert} < 0.7$. The error (in brackets) includes higher orders in $\als$ and in the multipole expansion, as well as  
the average of the linear term in the Cornell potential in order to account for the difference between weak and strong coupling regimes.}
	\label{decayW}
\end{table}	
	
Model independent results for hybrid decays in the Born-Oppenheimer approximation have been obtained before in \cite{Braaten:2014ita}. In that reference selection rules, based on the symmetries of the static limit, are obtained for a two-body decay of a hybrid to quarkonium plus a light meson, which constrain the possible quantum numbers of the latter. Our results are obtained under different assumptions, and may be regarded as complementary to those of ref. \cite{Braaten:2014ita}. First of all, our results hold beyond the static limit (e.g. $\Pi_u$-$\Sigma_u^-$ mixing is taken into account). Second, they are concerned with semi-inclusive decays, namely decays to quarkonium plus any state composed of light hadrons, rather than two-body decays. And third, they are based on the additional dynamical assumption that the decay process is short distance dominated. This assumption must be verified for each particular decay, and not always holds. In the cases it does, we are able to put forward not only constraints on quantum numbers (e.g. $L$ must be different from $J$ for a hybrid to decay to quarkonium) but also numerical estimates for the decay widths.

	\section{Mixing}\label{sec:mixing}
	
	So far we have not taken into account the possible mixing of hybrid states with other states that are known to exist in the same energy range, like quarkonium or heavy-light meson pairs,
 which may distort the spectrum and the decay properties. We shall focus here on the effects in the spectrum of the mixing with quarkonium, basically because they are amenable to a systematic treatment. 
In the static limit, quarkonium (the lowest  potential in fig. \ref{fig2}, $\Sigma_g^+$) and heavy hybrids (the remaining potentials in fig. \ref{fig2}) do not mix by construction (they are built as orthogonal states). 
Hence, the mixing must be due to $1/m_Q$ corrections to 
the Born-Oppenheimer approximation\footnote{In the weak coupling regime of pNRQCD, some ${\bf p}/m_Q$ contributions can be reshuffled into ${\bf r}i\partial_0$, which have the same size, by local field redefinitions \cite{Pineda:1997bj,Brambilla:1999xf,Brambilla:2004jw}. This is why singlet-octet transition terms appear in (\ref{pNRQCD'}) with no apparent $1/m_Q$ suppression.}. A way to systematically compute $1/m_Q$ corrections for quarkonium was established in \cite{hep-ph/0002250,hep-ph/0009145} for the strong coupling regime of pNRQCD, 
following earlier work in the literature \cite{Eichten:1980mw}. We show below how the formalism in \cite{hep-ph/0002250} can also be used to calculate the mixing potentials.
We may generally consider an effective theory for energy fluctuations $E$ around a hybrid state, such that $E\ll \lQ$. If there is
a heavy quarkonium state close to that energy, we may expect it to modify the value of the energy $E$. This effective theory reads, 	
\begin{equation}
	\begin{aligned}
	\label{H+S}
	\mathcal{L}_{H+S}=&
	   {\rm tr}\left( S^\dagger [i\partial_0-h_s]S\right) +\\ &+ {\rm tr}\left( H^{i\dagger}[i\delta_{ij}\partial_0-{h_H}_{ij}]H^j\right)+\\
		&+ {\rm tr}\left(S^\dagger  V_S^{ij} \left\{ \sigma^i \, , H^j\right\}+\,{\rm H.c.}\right) \,.
	\end{aligned}
	\end{equation}
	The traces are over spin indices and 
	\be
	\label{mixdecomp}
	V_S^{ij}=V_S^{ij}({\bf r})=\delta^{ij}V_S^\Pi(r) +{\hat r}^i{\hat r}^j(V_S^\Sigma(r)-V_S^\Pi(r)) \,,
	\ee
	is the mixing potential, $h_s=-\frac{\boldsymbol{\nabla^2}}{m_Q} + V_{\Sigma_g^+}(r)$ and ${h_H}_{ij}$ is defined in (\ref{h}). $S=S({\bf R},{\bf r}, t)$ transforms like {\bf H} under heavy quark spin symmetry 
and as follows under the discrete symmetries \cite{Brambilla:2004jw},
\bea
P: & \, S({\bf R},{\bf r}, t)& \rightarrow -S(-{\bf R},-{\bf r}, t)\nn\\
T: &\, S({\bf R},{\bf r}, t)& \rightarrow \sigma^2 S({\bf R},{\bf r}, -t)\sigma^2\\
C: &\,  S({\bf R},{\bf r}, t)& \rightarrow \sigma^2S^T({\bf R},-{\bf r}, t)\sigma^2  \,.\nn
\label{symS}
\eea
The transformations above together with those in (\ref{symH}) dictate the form of the last (mixing) term in (\ref{H+S}). The form of $V_S^{ij} ({\bf r})$ then follows from the symmetries of the static limit (see for instance \cite{Brambilla:2004jw}). Notice that in (\ref{H+S}) we only include the $1/m_Q$ corrections relevant to the mixing. There are also $1/m_Q$ corrections to $h_s$ \cite{hep-ph/0002250} and to ${h_H}_{ij}$, briefly discussed in Sec. \ref{discussion}, that we do not consider. For systems with the quark and antiquark of different flavor, two more terms are possible, which vanish in the equal mass limit,
\bea
\delta \mathcal{L}_{H+S}&=& {\rm tr}\left(S^\dagger  V_S'^{\, ij} \left[ \sigma^i \, , H^j\right]\right) +\\
&& + \, {\rm tr}\left(S^\dagger  V_L^{ij} L^i  H^j\right) + {\rm H.c.}\, ,\nn
\eea
where $L^i$ is the angular momentum operator.

\subsection{Matching to NRQCD at ${\cal O}(1/m_Q)$}

The NRQCD operators that create states at time $t$ with the same quantum numbers as $S$ and {\bf H} in the static limit read,
\bea
{\hat O}^\dagger({\bf r}, {\bf R}, t)&\equiv&\psi^\dagger (\frac{\bf r}{2},t)W( \frac{\bf r}{2}, -\frac{\bf r}{2}; t) \chi (-\frac{\bf r}{2},t) \nn \\
&=&Z^{1/2}_S(r) S^\dagger({\bf r}, {\bf R}, t)\\
{\hat O}_B^{\dagger i}({\bf r}, {\bf R}, t)&\equiv& \psi^\dagger (\frac{\bf r}{2},t)W( \frac{\bf r}{2}, {\bf R}; t) B^i({\bf R}; t) W( {\bf R}, -\frac{\bf r}{2}; t) \times \nn\\ 
&\times&\chi (-\frac{\bf r}{2},t)
= (Z^{1/2}_H)^{ij}({\bf r}) {H^\dagger}^j({\bf r}, {\bf R}, t) \,,\nn
\eea
where $W( {\bf r}, {\bf r}'; t)$ are straight Wilson lines joining the points {\bf r} and ${\bf r}'$ at a fixed time $t$, and 
\be
(Z^{1/2}_H)^{ij}({\bf r})=Z^{1/2}_\Sigma (r) {\hat r}^i{\hat r}^j+Z^{1/2}_\Pi(r)\left(\delta^{ij}-{\hat r}^i{\hat r}^j\right)
\, .
\ee
In the static limit we have,
\bea
&&<0\vert T\{{\hat O}^\dagger({\bf r}', {\bf R}', T/2){\hat O}({\bf r}, {\bf R}, -T/2)\}\vert 0> \nn\\
&&= <1>_\Box \delta ({\bf r}'-{\bf r})\delta ({\bf R}'-{\bf R})\\
&&<0\vert T\{ {\hat O}_B^{\dagger i}({\bf r}', {\bf R}', T/2){\hat O}_B^j({\bf r}, {\bf R}, -T/2)\}\vert 0> \nn\\
&&= <B^i({\bf R}',T/2)B^j({\bf R}; -T/2)>_\Box \delta ({\bf r}'-{\bf r})\delta ({\bf R}'-{\bf R})\, ,\nn
\eea
where $<\dots>_\Box$ means insertions in the square Wilson loop going from $-T/2$ to $T/2$ with spatial boundaries at ${\bf R}\pm {\bf r}/2$. In particular $<1>_\Box$ is the Wilson loop itself.
The matching calculation at ${\cal O}(1)$ leads to,
\bea
&&<1>_\Box =Z_S e^{-iTV_{\Sigma_g^+}(r)} \\
&&<B^i({\bf R},T/2)B^j({\bf R}; -T/2)>_\Box \nn\\
&&= {\hat r}^i{\hat r}^j Z_\Sigma e^{-iTV_{\Sigma_u^-}(r)} +\left(\delta^{ij}-{\hat r}^i{\hat r}^j\right)Z_\Pi e^{-iTV_{\Pi_u} (r)}\,.\nn
\eea
Hence $V_{\Sigma_u^-}(r)$ and $V_{\Pi_u}(r)$ can be obtained from large $T$ behavior of certain operator insertions in the Wilson loop, and
are known since long from lattice calculations \cite{Perantonis:1990dy,Bali:2000gf,Juge:2002br}.

The NRQCD Lagrangian density at  ${\cal O}(1/m_Q)$ reads
\bea
\mathcal{L}_{NRQCD}&=&
	\psi^\dagger\left[ iD_0+\frac{{\bf D}^2}{2m_Q}+gc_F \frac{\boldsymbol{\sigma}{\bf B}}{2m_Q}\right]\psi \nn\\
	&&+\chi^\dagger\left[ iD_0-\frac{{\bf D}^2}{2m_Q}-gc_F \frac{\boldsymbol{\sigma}{\bf B}}{2m_Q}\right] \chi  \, ,
	\label{NRQCD}
\eea
where $c_F$ is a matching coefficient that will eventually be approximated by its tree level value $c_F=1$.
Since the Lagrangian above contains a spin-dependent term, we expect the leading contribution to $V_S^{ij}$ to appear at ${\cal O}(1/m_Q)$. We can easily get it by
matching the following correlation function at ${\cal O}(1/m_Q)$,
\bea
&&<0\vert T\{{{\hat O}}({\bf r}', {\bf R}', T/2){{\hat O}}_B^{\dagger i}({\bf r}, {\bf R}, -T/2)\}\vert 0>\\
&&=Z_S^{1/2} (Z^{1/2}_H)^{ij}({\bf r}) <0\vert T\{S({\bf r}',{\bf R}', T/2){H^\dagger}^j({\bf r}, {\bf R}, -T/2)\}\vert 0>  \,\nn
\label{match}
\eea
and focusing on the spin-dependent terms. The lhs is calculated using first order in perturbation theory in $1/m_Q$ in NRQCD (\ref{NRQCD}). The rhs is calculated again at
first order in perturbation theory in $1/m_Q$ from (\ref{H+S}) (recall that $V_S^{ij}$ is treated as ${\cal O} (1/m_Q)$). Taking into account (\ref{mixdecomp}), we obtain,
\bea
\label{lattice}
&&\frac{\frac{g\,c_F}{2m_Q}\int_{-T/2}^{T/2} dt <{\bf \hat r}{\bf B}(\frac{\bf r}{2},t){\bf \hat r}{\bf B}({\bf 0},-T/2)>_\Box}
{<1>_\Box^{1/2}<{\bf \hat r}{\bf B}({\bf 0},T/2){\bf \hat r}{\bf B}({\bf 0},-T/2)>_\Box^{1/2}} \nn\\
&&=2V_S^\Sigma \frac{\sin \left( (V_{\Sigma_u^-}-V_{\Sigma_g^+})T/2\right)}{ V_{\Sigma_u^-}-V_{\Sigma_g^+}}\\
&&\frac{\frac{g\,c_F}{2m_Q}\int_{-T/2}^{T/2} dt}{<1>_\Box^{1/2}}\times \nn\\
&&\frac{ <{\bf B}(\frac{\bf r}{2},t){\bf B}({\bf 0},-T/2)-{\bf \hat r}{\bf B}(\frac{\bf r}{2},t){\bf \hat r}{\bf B}({\bf 0},-T/2)>_\Box}
{<{\bf B}({\bf 0},T/2){\bf B}({\bf 0},-T/2)-{\bf \hat r}{\bf B}({\bf 0},T/2){\bf \hat r}{\bf B}({\bf 0},-T/2)>_\Box^{1/2}} \nn\\
&&=2\sqrt{2}V_S^\Pi \frac{\sin \left( (V_{\Pi_u}-V_{\Sigma_g^+})T/2\right)}{ V_{\Pi_u}-V_{\Sigma_g^+}}\,.
\eea
Notice that the Euclidean version of the objects on the lhs can be easily calculated on the lattice. At large $T$, $V_S^\Sigma$ and $V_S^\Pi$ can be then extracted by matching the data to the Euclidean version of the rhs, 
once $V_{\Sigma_g^+}$, $V_{\Sigma_u^-}$ and $V_{\Pi_u}$ are known. In the following sections we are going to derive short and long distance constraints on these potentials using weak coupling pNRQCD \cite{Pineda:1997bj,Brambilla:1999xf} and the QCD 
effective string theory \cite{Luscher:2002qv,Luscher:2004ib} respectively.

\subsubsection{Short dinstance constraints}

At short distances, the time evolution of a $Q\bar Q$ pair is described by the weak coupling regime of pNRQCD \cite{Pineda:1997bj,Brambilla:1999xf}, 
the Lagrangian of wich has been displayed in (\ref{pNRQCD'}) at next-to-leading order in the multipole expansion. The operators ${\hat O}({\bf r}, {\bf R}, t)$ and ${\hat O}_B^i({\bf r}, {\bf R}, t)$ match onto the singlet field
$S({\bf r}, {\bf R}, t)$ and the operator ${\rm tr} (O({\bf r}, {\bf R}, t)B^i({\bf R}, t))$ respectively. The leading spin-dependent term in the pNRQCD Lagrangian reads \cite{Brambilla:2002nu},
\bea
\label{pNRQCD''}
	{\mathcal{L}'}_{pNRQCD}&=&{gc_F \over 2m_Q} {\rm Tr} \left( {\rm O}^\dagger({\bf r}, {\bf R}, t)    {\bf B}({\bf R}, t)\,\{ \boldsymbol{\sigma} , {\rm S}({\bf r}, {\bf R}, t)\} \right)\nn \\
	+ \hbox{H.c.}\, .
	\eea
We use tr for trace over color indices and Tr for trace over both color and spin indices. Notice that the term above shows an
{\bf r}-independent interaction between	the singlet field and the operator ${\rm tr} (O\,B^i)$, which implies that
\be
V_S^\Sigma(r)=V_S^\Pi(r)= \pm\frac{c_F \lambda^2}{m_Q}\, ,
\label{sdc}
\ee
where  $\lambda\sim \lQ$ is a constant, and we have put the sign explicitly.

\subsubsection{Long dinstance constraints}

At long distances the energy spectrum of a static $Q\bar Q$ pair is well described by the QCD effective string theory (EST) \cite{Luscher:2002qv,Luscher:2004ib}. The mapping between operator insertions in the temporal Wilson lines of the 
Wilson loop and the corresponding operators in the EST was established in \cite{PerezNadal:2008vm}, following earlier work \cite{Kogut:1981gm}. For the relevant operators to us, it reads,

\bea
	\label{simetries}
	&&
	B^l(t,{\bf r}/2)\rightarrow \Lambda'\epsilon^{lm}\partial_t\partial_z\xi^m(t,r/2)\nn\\
	&&
	B^l(t,-{\bf r}/2) \rightarrow \Lambda'\epsilon^{lm}\partial_t\partial_z\xi^m(t,-r/2)\\
	&&
	B^3(t,{\bf r}/2)\rightarrow \Lambda'''\epsilon^{lm}\partial_t\partial_z\xi^l(t,r/2)\partial_z\xi^m(t,r/2)\nn\\
	&&
	B^3(t,-{\bf r}/2)\rightarrow \Lambda'''\epsilon^{lm}\partial_t\partial_z\xi^l(t,-r/2)\partial_z\xi^m(t,-r/2)\,,\nn
	\eea
where $l\,,m =1\,,2$.    
Here, we also need to map the states created by operator insertions in spacial Wilson lines to the corresponding states in EST. In order to do so it is convenient to take the {\bf r} along the $z$ axis, and write the EST Lagrangian in terms
of the complex field $\varphi (z,t)=(\xi^1(z,t)+i\xi^2(z,t))/\sqrt{2}$. This field has nice transformation properties under $D_{\infty h}$, the relevant space group,
\bea
&R_z (\theta)&: 	\varphi (z,t)\rightarrow e^{i\theta}\varphi (z,t)\nn\\
&P &: \varphi (z,t) \rightarrow -\varphi (-z,t)\\
&R_{xz} &: \varphi (z,t)\rightarrow \varphi^\ast (z,t)\nn \, ,
\eea
where $R_z(\theta)$, $P$ and $R_{xz}$ stand for a rotation of angle $\theta$ around the $z$ axis, a parity transformation, and a reflexion through the $xz$ plain respectively. The Lagrangian density at LO reads,
\be
{\cal L}_{EST}=\kappa \partial_\mu \varphi \partial^\mu \varphi^\ast \, ,
\ee
where $\kappa$ is the string tension and $ \varphi (z,t)$ fulfils Dirichlet boundary conditions, $ \varphi (r/2,t)=\varphi (-r/2,t)=0$. $ \varphi (z,t)$ can written in terms of creation and annihilation opertators
\bea
 && \varphi (z,t)=\sum_{n=1}^\infty\frac{1}{2E_n}\left( e^{-iE_nt}\varphi_n(z) a_n 
+ e^{iE_nt}\varphi_n^\ast (z) b_n^\dagger \right) \nn\\
&& \varphi_n(z)=\frac{1}{\sqrt{2r}}\left(e^{iE_nz}+(-1)^{n+1}e^{-iE_nz}\right)\\
&& [a_n,a_m^\dagger]=[b_n,b_m^\dagger]=\frac{2E_n}{\kappa}\delta_{nm} \quad , E_n=\frac{\pi n}{r} \, .\nn
\eea 
The remaining commutators vanish.
$a_n^\dagger$ ($b_n^\dagger$) on the vacuum creates a state of energy $E_n$, angular momentun $1$ ($-1$) and parity $(-1)^n$. The reflexion with respect the $xz$ plain interchanges $a_n \leftrightarrow b_n$. If we define
\bea
{\hat O}_B^\dagger ({\bf r}, {\bf 0}, t)&=&{\hat O}_B^{\dagger 1}({\bf r}, {\bf 0}, t)+i{\hat O}_B^{\dagger 2}({\bf r}, {\bf 0}, t)\\
{\hat O}_B^{\dagger \ast}({\bf r}, {\bf 0}, t)&=&{\hat O}_B^{\dagger 1}({\bf r}, {\bf 0}, t)-i{\hat O}_B^{\dagger 2}({\bf r}, {\bf 0}, t) \, ,\nn
\eea
then the following identifications fulfil the $D_{\infty h}$ symmetry requirements,
\bea
\label{statesmap}
&{\hat O}^\dagger ({\bf r}, {\bf 0}, -T/2)\vert 0> &\rightarrow  \vert 0>\nn\\
&{\hat O}_B^{\dagger 3} ({\bf r}, {\bf 0}, -T/2)\vert 0>&\rightarrow \frac{\kappa}{2\sqrt{2E_1E_2}}\left(a_1^\dagger b_2^\dagger -b_1^\dagger a_2^\dagger \right)\vert 0>\nn\\
&{\hat O}_B^{\dagger } ({\bf r}, {\bf 0}, -T/2)\vert 0>&\rightarrow \sqrt{\frac{\kappa}{2 E_1}} b_1^\dagger \vert 0>\\
&{\hat O}_B^{\dagger \ast} ({\bf r}, {\bf 0}, -T/2)\vert 0>&\rightarrow -\sqrt{\frac{\kappa}{2 E_1}} a_1^\dagger \vert 0>\, .\nn
\eea
Let us perform analogous definitions for the chromomagnetic fields,
\bea 
B(t,{\bf r})&=&B^1(t,{\bf r})+iB^2(t,{\bf r})\\
B^\ast(t,{\bf r})&=&B^1(t,{\bf r})-iB^2(t,{\bf r})\, .\nn
\eea
The mapping (\ref{simetries}) then reads,
\bea
B(t,\pm{\bf r}/2)&\rightarrow&-i\sqrt{2}\Lambda'\partial_t\partial_z \varphi (\pm r/2, t)\nn\\
B^\ast(t,\pm{\bf r}/2)&\rightarrow&i\sqrt{2}\Lambda'\partial_t\partial_z \varphi^\ast (\pm r/2, t)\\
B^3(t,\pm{\bf r}/2)&\rightarrow&i\Lambda'''\partial_t\partial_z \varphi (\pm r/2, t)\partial_z \varphi^\ast (\pm r/2, t)+{\rm H.c.}\, .\nn
\eea
Upon substituting the above expressions and (\ref{statesmap}) in (\ref{lattice}), we obtain,
\be
\label{ldc}
V_S^\Sigma(r)=-\frac{\pi^2g\Lambda'''c_F}{m_Q\kappa r^3}\quad ,\quad V_S^\Pi(r)=\frac{\pi^{3/2}g\Lambda'c_F}{2m_Q\sqrt{\kappa} r^2}\, .
\ee
The parameters $g\Lambda'\sim \lQ$ and $g\Lambda'''\sim \lQ$ also appear in the spin-orbit  and tensor potentials of heavy quarkonium \cite{PerezNadal:2008vm,Brambilla:2014eaa} , 
which have been calculated on the lattice \cite{Bali:2000gf,Koma:2006fw}. We obtain from fits to the data of ref. \cite{Koma:2009ws},
\be 
g\Lambda'\sim -59 {\rm MeV} \quad ,\quad g\Lambda'''\sim \pm 230 {\rm MeV}\,,
\ee
Details on the fits are given in the Appendix \ref{fits}\footnote{Due to an error in the identification, the value of $g\Lambda'$ displayed in \cite{Oncala:2016wlm} as $\Lambda'$ is twice the actual value. This change does not affect the statements made in that paper.}.

\subsubsection{Modeling the mixing potential} 

For the actual mixing potentials we use a simple interpolation between (\ref{sdc}) and (\ref{ldc}) that allows for a sign flip between the short and long distance expressions without introducing any further scale, namely,

\bea
V_S^\Pi[\pm -](r)&=&\frac{\lambda^2}{m_Q}\left(\frac{\pm 1-(\frac{r}{r_\Pi})^2}{1+(\frac{r}{r_\Pi})^4}\right) \\
V_S^\Sigma[\pm\pm](r)&=&\frac{\lambda^2}{m_Q}\left(\frac{\pm 1\pm (\frac{r}{r_\Sigma})^2}{1+(\frac{r}{r_\Sigma})^5}\right) ,
\eea
where $r_\Pi=(\frac{\vert g\Lambda'\vert\pi^{\frac{3}{2}}}{2\lambda^2\kappa^{\frac{1}{2}}})^{\frac{1}{2}}$ and $r_\Sigma=(\frac{\vert g\Lambda'''\vert\pi^2}{\lambda^2\kappa})^{\frac{1}{3}}$. Note that the $\pm$ in $V_S^\Pi$ and the first $\pm$ in $V_S^\Sigma$  are correlated because both potentials have the same short distance behavior.
 We will explore the following values for the only unknown parameter $\lambda$, $\lambda= 100, 300, 600$ MeV, and all possible sign combinations for the $1^{--}$ charmonium states below, and choose the one that suits better the phenomenology. 

\subsection{Mixing equations}

Now, we need to include the quark spin degree of freedom to the equations displayed in section II. Let us write,
\bea
\label{spindecomp}
S&=&\frac{1}{\sqrt{2}}\left(S_0+\sigma^kS_1^k\right)\\
H^j&=&\frac{1}{\sqrt{2}}\left(H^j_0+\sigma^iH_1^{ji}\right)\, , \nn
\eea
where we have omited the arguments in $S=S({\bf R},{\bf r}, t)$ and $H^j=H^j({\bf R},{\bf r}, t)$, the subscript $0,1$ stands for the total spin of the quark-antiquark pair, and the superscripts $k,i$ label the three states in the spin $1$ case. Recall that the superscript $j$ labels the three states of the total angular momentum $1$ of the gluonic degrees of freedom. Then the last term in (\ref{H+S}) reads,
\be
{\rm tr}\left(S^\dagger  V_S^{ij} \left\{ \sigma^i \, , H^j\right\}\right)=2V_S^{ij}\left({S_1^i}^\dagger H_0^j+{S_0}^\dagger H_1^{ji} \right)\, .
\ee
Note that this term mixes spin $0$ ($1$) hybrids with spin $1$ ($0$) quarkonium.
In view of the decomposition of $V_S^{ij}$ in (\ref{mixdecomp}), we only need to analyze ${S_1^j}^\dagger H_0^j$, ${S_1^i}^\dagger H_0^j {\hat r}^i{\hat r}^j$, ${S_0}^\dagger H_1^{jj}$ and ${S_0}^\dagger H_1^{ji} {\hat r}^i{\hat r}^j$. Consider the first expression,
\be
\int d\Omega {S_1^j}^\dagger H_0^j=\sum_{JML} \frac{S_{1JM}^{L \dagger}P_{0JM}^L}{r^2}=\sum_{\j\m L} \frac{S_{1\j\m}^{L \dagger}P_{0\j\m}^L}{r^2}\, ,
\ee
where $J$ is the orbital angular momentum, plus the quark spin for $S_1$ and plus the gluonic total angular momentum for $H_0$, and $M$ its third component. $J=\j$ and $M=\m$, the total angular momentum and its third component respectively. $L=J,J \pm 1$ are simply denoted by $0, \pm $. We have used above the same decomposition for $S_1^j$ as the one used for $H^j$ in (\ref{jJML}). For the second expression we have,
\bea
&& \int d\Omega {S_1^i}^\dagger H_0^j {\hat r}^i{\hat r}^j= \\
&& \frac{1}{r^2} S_{1\j\m}^{- \dagger}\left( \frac{\j}{2\j+1}P_{0\j\m}^- -\frac{\sqrt{\j(\j+1)}}{2\j+1}P_{0\j\m}^+\right)\nn\\
&& + \frac{1}{r^2} S_{1\j\m}^{+ \dagger}\left( \frac{\j+1}{2\j+1}P_{0\j\m}^+ -\frac{\sqrt{\j(\j+1)}}{2\j+1}P_{0\j\m}^-\right)
\, .\nn
\eea
For the third and fourth expressions, we need to introduce tensor spherical harmonics $Y^{ji\, LJ}_{\j\m}({\hat \bf r})$ (see appendix \ref{tensor}), which are eigenfunctions of {\bf S}$^2$, {\bf L}$_g^2$, {\bf L}$^2$, {\bf J}$^2$, ${\pmb{\j}}^2$ and $\j_3$ with eigenvalue $2$, $2$, $L(L+1)$, $J(J+1)$, $\j(\j+1)$ and $\m$ respectively, with $L=J,J \pm 1$, $J=\j,\j\pm 1$ . We then have,
\be
H^{ji}_1({\bf r},t)=\frac{1}{r}\sum_{LJ\j\m} Y^{ji\, LJ}_{\j\m}({\hat {\bf r}}) P^{LJ}_{1\j\m}(r) \, .
\ee
We will use $0,\pm$ both for $L=J,J\pm 1$ and $J=\j,\j \pm 1$.
Hence,
\bea
&&\int d\Omega {S_0}^\dagger H_1^{jj}=
\frac{1}{r^2}\sum_{\j\m} S^\dagger_{0\, \j\m}\times \\
&&\left( -\sqrt{\frac{2\j-1}{2\j+1}}P_{1\j\m}^{+ -}+P_{1\j\m}^{0 0}
-\sqrt{\frac{2\j+3}{2\j+1}}P_{1\j\m}^{- +}\right) \nn
\eea

\bea
&& \int d\Omega {S_0}^\dagger H_1^{ji} {\hat r}^i{\hat r}^j=
\frac{1}{r^2}\sum_{\j\m} S^\dagger_{0\, \j\m}\times \\
&&\left( \sqrt{\frac{\j(\j-1)}{(2\j+1)(2\j-1)}}P_{1\j\m}^{- -}
-\frac{\j}{\sqrt{(2\j+3)(2\j+1)}}P_{1\j\m}^{+ -}+\right.\nn\\
&&\left. + \sqrt{\frac{(\j+1)(\j+2)}{(2\j+1)(2\j+3)}}P_{1\j\m}^{+ +} 
-\frac{\j+1}{\sqrt{(2\j+3)(2\j+1)}}P_{1\j\m}^{- +}\right) \nn
\,.
\eea

Putting all together, we get the following sets of coupled equations. For $S=0$ hybrids, we have 
for $\j\not= 0$
\bea
\label{cop}
&& 	\left[ 
 -\frac{1}{m_Q}\frac{\partial^2}{\partial r^2}\!
 + \frac{\j(\j+1)}{m_Qr^2}    
	+ \begin{pmatrix}
 		\!V_{\Sigma_g^+}  & 2V_S^\Pi                    \\
 		2V_S^\Pi & 	V_{\Pi_u}                   \\
 	\end{pmatrix}
 	\!
 	\right]\! 
	\begin{pmatrix}
 		S_{1\, \j\m}^{0}(r) \\
 		P_{0\, \j\m}^{0}(r) \\
 	\end{pmatrix}\!
	\nn\\
	&&
 	=\!E\!
 	\begin{pmatrix} 
 	S_{1\, \j\m}^{0}(r) \\
 		P_{0\, \j\m}^{0}(r) \\		
 	\end{pmatrix} \,.
	\eea

	\begin{widetext}
{
	\scriptsize	
\be
	\label{cop+-}
 	\left[ 
 -\frac{1}{m_Q}\frac{\partial^2}{\partial r^2}\! +
\begin{pmatrix}
 \frac{(\j+1)(\j+2)}{m_Qr^2}+V_{\Sigma_g^+}    & 0 & 2(V_S^\Pi+\frac{\j+1}{2\j+1}V_S^q) & -2 V_S^q\frac{\sqrt{\j(\j+1)}}
{2\j+1}\\
0 & \frac{(\j-1)\j}{m_Qr^2}+V_{\Sigma_g^+} & -2 V_S^q\frac{\sqrt{\j(\j+1)}}{2\j+1} & 2(V_S^\Pi+\frac{\j}{2\j+1}V_S^q ) \\
2(V_S^\Pi+\frac{\j+1}{2\j+1}V_S^q) & -2 \frac{\sqrt{\j(\j+1)}}{2\j+1} & \frac{(\j+1)(\j+2)}{m_Qr^2}+V_{\Sigma_u^-}+\frac{\j}{2\j+1}V_q & \frac{\sqrt{\j(\j+1)}}
{2\j+1} V_q \\
-2 V_S^q\frac{\sqrt{\j(\j+1)}}{2\j+1} & 2(V_S^\Pi+\frac{\j}{2\j+1}V_S^q) & \frac{\sqrt{\j(\j+1)}}
{2\j+1} V_q & \frac{(\j-1)\j}{m_Qr^2}+V_{\Sigma_u^-}+\frac{\j+1}{2\j+1}V_q \\
 	\end{pmatrix}
 	\!
 	-E\right]\! 
	\begin{pmatrix}
 		S_{1\, \j\m}^{+}(r) \\ S_{1\, \j\m}^{-}(r) \\
 		P_{0\, \j\m}^{+}(r) \\ P_{0\, \j\m}^{-}(r) \\
 	\end{pmatrix}\!
 	= 0\,,
	\ee
	}
	where $V_S^q=V_S^\Sigma-V_S^\Pi$.
	For $\j=0$, equations (\ref{cop}) do not exist and equations (\ref{cop+-}) reduce to two 
coupled equations for $S_{1\, 00}^{+}(r)$ and $P_{0\, 00}^{+}(r)$. For $S=1$ hybrids, 
$ P_{1\, \j\m}^{+0}(r)$ and $ P_{1\, \j\m}^{-0}(r)$ do not couple to heavy quarkonium. The remaining 
do it according to the following equations for $\j > 1$,
	
	{
\scriptsize

\bea
	\label{cop1}
 &&	\left[ 
 \frac{1}{m_Q}\frac{\partial^2}{\partial r^2}\! -
\begin{pmatrix}
 \frac{\j(\j+1)}{m_Qr^2}+V_{\Sigma_g^+}&  V_S^{++} & V_S^{-+}& V_S^{+-} &  V_S^{--} & V_S^\Pi\\
V_S^{++} & \frac{(\j+2)(\j+3)}{m_Qr^2}+V^{++}& -2 V_q\frac{\sqrt{(\j+1)(\j+2)}}{2\j+3} & 0 & 0 & 0 \\
V_S^{-+} & -2 V_q\frac{\sqrt{(\j+1)(\j+2)}}{2\j+3}  & \frac{\j(\j+1)}{m_Qr^2}+V^{-+}& 0 & 0 & 0 \\
V_S^{+-} & 0 & 0 & \frac{\j(\j+1)}{m_Qr^2}+V^{+-}&  V_q\frac{\sqrt{(\j-1)\j}}{2\j-1} & 0 \\
V_S^{--} & 0 & 0 & V_q\frac{\sqrt{(\j-1)\j}}{2\j-1} & \frac{(\j-1)\j}{m_Qr^2}+V^{--}& 0 \\
V_S^\Pi & 0 & 0 & 0 & 0 & \frac{\j(\j+1)}{m_Qr^2}+V_{\Pi_u}\\
 	\end{pmatrix} +E 
	\right]\! 
	\begin{pmatrix}
 		S_{0\, \j\m}(r) \\ P_{1\, \j\m}^{++}(r) \\
 		P_{1\, \j\m}^{-+}(r) \\ P_{1\, \j\m}^{+-}(r) \\
P_{1\, \j\m}^{--}(r)\\ P_{1\, \j\m}^{00}(r)
 	\end{pmatrix}\!=0\,,
	\nn \\
	&& 
	\eea
	}
	\end{widetext}

	where,
	\bea 
	V_S^{++} & = & \sqrt{\frac{(\j+1)(\j+2)}{(2\j+1)(2\j+3)}}V_S^q \nn\\
	V_S^{-+} & = & -\frac{\j+1}{\sqrt{(2\j+3)(2\j+1)}}V_S^q-\sqrt{\frac{2\j+3}{2\j+1}}V_S^\Pi\nn\\
	V_S^{+-} & = & -\frac{\j}{\sqrt{(2\j+3)(2\j-1)}}V_S^q -\nn \\
	         & & - V_S^\Pi\frac{\sqrt{(2\j-1)(2\j+1)}}{2\j+1}\nn
\eea

\bea					
	V_S^{--} & = & \sqrt{\frac{\j(\j-1)}{(2\j+1)(2\j-1)}}V_S^q \nn \\
	V^{++} & = & V_{\Sigma_u^-}+\frac{\j+1}{2\j+3}V_q \nn \\
	V^{-+} & = & V_{\Sigma_u^-}+\frac{\j+2}{2\j+3}V_q \\
	V^{+-} & = & V_{\Sigma_u^-}+\frac{\j-1}{2\j-1}V_q \nn\\
	V^{--} & = & V_{\Sigma_u^-}+\frac{\j}{2\j-1}V_q \,.\nn
	\eea
	
For $\j=0$, $P_{1\, 00}^{00}(r)$, $P_{1\, 00}^{--}(r)$ and $P_{1\, 00}^{+-}(r)$ do not exist, and the system 
reduces to the three upper equations. $P_{1\, 00}^{0-}(r)$, which does not couple to heavy quarkonium,
 does not exists either. For $\j=1$, $P_{1\, 1\m}^{--}(r)$ does not exists and the system above reduces 
to five coupled equations.

\subsection{Spectrum}
\label{sp}

In order to fix the signs and the parameter $\lambda$ of the mixing potentials, we focus on the spin zero $n(s/d)_1$ ($n=1,2,3$), states in table \ref{cEspectrum}, which can be identified with $Y(4008)$, $Y(4360)$ and $Y(4660)$. The main problem with this identification is that all three states have been observed to decay to spin one quarkonium states, which violates spin symmetry. However, according to eq. (\ref{cop+-}) the spin zero hybrids mix with spin one quarkonium, and hence, if this mixing is large, we may find a natural explanation to these decays. We present our results in table \ref{c1--} (the case $\lambda=100$ MeV is not displayed, it produces a tiny mixing in all cases). We observed that the case that provides the largest amount of mixing is the combination $V_S^\Pi [+-]$ with $V_S^\Sigma [++]$ and $\lambda=600$ MeV. This is the sign combination and the value of $\lambda$ that we will take for the rest of the paper. The spectrum of charmonium and charmonium hybrids is given in tables \ref{c1--}-\ref{c2-+} and the one of bottomonium and bottomonium hybrids in tables \ref{b1--}-\ref{b2-+}.
The general trend (with a few exceptions) is that hybrid states get heavier whereas quarkonium states get lighter due the mixing.

Since we have used leading order potential for both quarkonium and hybrids, the potentials we missed start at order $1/m_Q$. Hence the error to assign to this calculation for the hybrids is $\lQ^2/m_Q$, since $\lQ$ is the next relevant scale. For quarkonium this is not always the case, since the typical momenta can be larger than $\lQ$. A detailed error analysis is carried out in the Appendix \ref{quarkonium}. For simplicity, we will stick to the $\lQ^2/m_Q$ estimate for quarkonium as well. Taking $\lQ\sim 400$ MeV, we obtain a precision of about $110$ MeV for charmonium, and $33$ MeV for bottomonium. These are the numbers we will have in mind when comparing to experiment and to other approaches.

\begin{widetext}

\begin{table}[htbp]
	\centering
	
	\begin{tabular}{|c|c|c|c|c|c|c|c|c|c|c|c|c|c|c|c|c|c|c|}
		\cline{4-19}
		\multicolumn{3}{c}{}\hfill\vline & \multicolumn{4}{c}{$V_S^\Pi [+-]\,,V_S^\Sigma [++]$}\vline & \multicolumn{4}{c}{$V_S^\Pi [+-]\,,V_S^\Sigma [+-]$}\vline  & \multicolumn{4}{c}{$V_S^\Pi [--]\,,V_S^\Sigma [-+]$}\vline & \multicolumn{4}{c}{$V_S^\Pi [--]\,,V_S^\Sigma [--]$}\vline \\ \hline
		$NL_J$ & $\lambda=0$ & \%   & $\lambda=0.3$  &\%   & $\lambda=0.6$ &\%  & $\lambda=0.3$  &\%   & $\lambda=0.6$ &\% & $\lambda=0.3$  &\%   & $\lambda=0.6$ &\% & $\lambda=0.3$  &\%   & $\lambda=0.6$ &\%
		\\ \hline \hline
		$1s$ & 3.068 & 0 & 3.064 & 0 & 3.001 & 4 & 3.066 & 0 & 3.053& 0 & 3.063 & 0 & 3.036&2&3.061&1&2.989&6\\ 
		$2s$ & 3.678 & 0 & 3.672 & 1 & 3.628 & 14& 3.677 & 1 & 3.670 & 4 & 3.677 & 0&3.661 &4&3.672&1&3.630&7\\
		$1d$ & 3.793 & 0 & 3.773 & 4 & 3.687 & 12& 3.790 & 1 & 3.785 & 2 & 3.792 & 0& 3.789&0&3.782&1&3.712&7\\ 
		$1(s/d)_1$ &  4.011 & 100 & 4.016  & 96 & 4.014 & 71& 4.012 & 99 & 4.004& 96 & 4.014 & 99& 4.025&99&4.016&98&4.040&85\\
		$3s$ & 4.131 & 0 & 4.127 & 0 & 4.107 & 10& 4.128 & 1 & 4.130  & 7& 4.130 & 0 & 4.125&10&4.128&2&4.103&12\\  
		$2d$ & 4.210 & 0 & 4.203 & 20 & 4.180 & 79& 4.209 & 10 & 4.207 & 39 & 4.209 & 2& 4.205&5&4.204&1&4.172&52\\ 
		$2(s/d)_1$ & 4.355 & 100 & 4.358 & 97 & 4.366 & 65& 4.356 &  98& 4.355 & 89 & 4.357 & 100& 4.368&94&4.357&100&4.383&86\\ 
		$4s$ & 4.512 & 0 & 4.515 & 0 & 4.497 & 0& 4.517 & 1& 4.513 &  7& 4.517 &  0& 4.508&8&4.515&1&4.495&0\\
		$3d$ & 4.579 & 0 & 4.573 & 2 & 4.559 & 8& 4.578 & 0 & 4.574 & 5 & 4.578 & 1& 4.568&7&4.574&0&4.550&3\\
		$3(s/d)_1$ & 4.692 & 100 & 4.699 & 98 & 4.711 & 83& 4.694 & 99 & 4.699& 93 & 4.693 & 100& 4.699&97&4.698&99&4.724&90\\ 
		$4(s/d)_1$ & 4.718 & 100 & 4.730 & 100 & 4.785 & 96& 4.719 & 100 & 4.718& 98 & 4.720 & 100& 4.728&98&4.728&100&4.779&97\\ 
        $5s$ & 4.865	&	 0  & 4.864 & 0 & 4.848 & 3 & 4.865 &  0 & 4.865 & 7 & 4.865 & 0& 4.867&7&4.864&1&4.846&2\\ 
		$4d$ & 4.916 & 0& 4.913 & 7 & 4.903& 35 & 4.915 & 2 & 4.915&19&4.915& 0 &4.912&12 & 4.913 &3 &4.894 & 21\\ 
		$5(s/d)_1$ & 5.043 & 100 & 5.044&99 & 5.046 &84 & 5.043& 99&5.043 & 94&5.044 & 100&5.050 &97 & 5.044& 100& 5.067& 93\\ 
		 \hline
	\end{tabular}
	\caption{Spectrum of charmonium ($S=1$) and charmonium hybrids ($S=0$): $1^{--}$ states. Masses are in GeV. The \% columns show the fraction of the hybrid components for the mass states in the previous column. 
	$m_c=1.47$ GeV. }
	
	\label{c1--}
\end{table}

\end{widetext}

\section{Comparison with other approaches}
\label{sec:th}

In this section, we compare our results with other QCD based approaches. For convenience we will compare our results for the spectrum in the case $\lambda=0$ (no mixing). The shifts in the spectrum due to mixing are 
within our estimated errors.

\subsection{Born-Oppenheimer approximation}

In \cite{Juge:1999ie}, the lower lying bottomonium hybrid spectrum was calculated from the static potentials $\Pi_u$ and $\Sigma_u^-$ and normalized to the bottomonium spectrum. 
The mixing between hybrid states built out of these potentials that appears 
at leading order due to the kinetic term of the heavy quarks was ignored. The masses obtained for $H_1$ ($1(s/d)_1$), $H_2$ ($1p_1$), $H_3$ ($1p_0$), and $H_1'$ ($2(s/d)_1$) are between $150$-$300$ MeV heavier than ours. This is probably due to the different choice of the bottom quark mass.

In \cite{Braaten:2014qka}, the lower lying hybrid spectrum was calculated as above. However, for charmonium, 
the ground state for each potential was fixed to the lattice data of ref. \cite{Liu:2012ze}. The mixing between hybrid states was also ignored. If we compare the splittings obtained from table X of \cite{Braaten:2014qka} with those obtained from our tables \ref{cEspectrum} and \ref{bEspectrum},
 we find agreement within $20$ MeV, except for the $H_4$-$H_1$ case 
for which we obtain a lower value by about $40$ MeV and the $H_3'-H_3$ case for which we obtain a higher value of about $70$ MeV. We have identified the states $H_1$, $H_1'$, $H_2$, $H_2'$, $H_2''$, $H_3(1P)$ and $H_4$  with $1(s/d)_1$, $2(s/d)_1$, $1p_1$, $1d_2$, $2p_1$, $3(s/d)_1$ and $1(p/f)_2$ 
respectively.

Our hybrid spectrum is compatible within errors with that of ref. \cite{Berwein:2015vca} both for charmonium and bottomonium, except for the bottomonium $1(s/d)_1$ and $2p_0$ states, 
for which we have slightly lower masses. Our central values tend to be at the lower end of their error bars. 
Although the construction of the effective theory for hybrids is somewhat different and the parametrization of the potentials as well, the most relevant difference is probably 
the normalization of the spectrum. Indeed, in ref. \cite{Berwein:2015vca} the hybrid spectrum is normalized using the charm and bottom masses in the RS scheme \cite{hep-ph/0105008}, whereas here we normalize it to 
the corresponding quarkonium spectrum, which is not calculated in that reference. We have checked that we reproduce the results of ref. \cite{Berwein:2015vca} with our code if we input their 
potentials\footnote{We have also checked that our results are reproduced by the code of ref. \cite{Berwein:2015vca} if our potentials are input. We thank the authors of that reference for 
providing their code for the test.}.

\subsection{Lattice QCD}

In \cite{hep-lat/0210030}, the spectrum of the lightest exotic charmonium hybrids is calculated in the quenched approximation for a relativistic charm action in an anisotropic lattice ($a_s=
0.197-0.09$ fm, $a_s/a_t$=2). Their results for the $1^{-+}$, $0^{+-}$ and $2^{+-}$ states are between $400-700$ MeV higher than ours.

There has been a recent update \cite{Cheung:2016bym} of earlier results \cite{Liu:2012ze} by the Hadronic Spectrum Collaboration for the charmonium spectrum including hybrid states. They use relativistic charm and dynamical light quarks in an anisotropic lattice with temporal spacing $a_t \sim 0.034$ fm and spatial spacing $a_s \sim 0.12$ fm. The update basically consists of taking up and down quark masses smaller than in the previous calculation ($m_\pi\sim 240$ MeV and $m_\pi\sim 400$ MeV respectively). The hierarchy of the lowest lying hybrid multiplets agrees with ours, from lighter to heavier: $1(s/d)_1$,  $1p_1$, $1(p/f)_2$ and $1p_0$. However, their numbers are considerable larger than ours: $381$ MeV, $326$ MeV, $392$ MeV and $151$ MeV higher for the spin average of the $1(s/d)_1$,  $1p_1$, $1(p/f)_2$ and $1p_0$ multiplets respectively. The hierarchy in which quarkonium and hybrid states arise agrees for the $1^{++}$ (4 states) and $1^{+-}$ (6 states) quantum numbers but disagrees for the remaining non-exotic ones. 

In \cite{1603.06467}, the lower lying charmonium spectrum is also calculated with four dynamical quarks in a Wilson twisted mass action. Lattice spacings ranging from 
$0
.
0619$ fm to 
$0.
0885$ fm and pion masses ranging from $225$MeV to $470$ MeV are used and both the continuum and the chiral extrapolations are carried out. They find a $1^{--}$ state at $3951$ MeV that is compatible with our $1(s/d)_1$ spin zero hybrid state ($4011$ MeV). With less significance, they also find two $2^{++}$ states at about $4460$ MeV and $4530$ MeV which are compatible with our $2f$ quarkonium ($4428$ MeV) and $2(p/f)_2$ spin zero hybrid ($4563$ MeV) respectively.

In \cite{Juge:1999ie}, the bottomonium hybrid spectrum is calculated in quenched lattice NRQCD  using an anisotropic lattice ($a_s \sim 0.11$ fm, $a_s/a_t = 3$). They find the lightest hybrid $H_1$ ($1(s/d)_1$) $1.49(2)(5)$ GeV above the $1S$ quarkonium, this is about 
$250$ MeV heavier than ours. About the same difference is also found for $H_2$ ($1p_1$) and $H_3$ ($1p_0$), whereas for $H_1'$ ($2(s/d)_1$) the difference raises to $470$ MeV.

For bottomonium, there is also a quenched lattice calculation with relativistic bottom quarks in an anisotropic lattice ($a_s \sim 0.04-0.17$ fm, $a_s/a_t=4,\, 5$) \cite{hep-lat/0111049}. 
The masses for the  lightest $2^{--}$, $1^{-+}$ and $2^{+-}$ hybrids are displayed, which turn out to be either lighter ($2^{--}$) or heavier ($1^{-+}$ and $2^{+-}$) than our results, 
in spite of the large errors ($200-600$ MeV).

\subsection{QCD sum rules}  

In \cite{1304.4522}, the hybrid spectrum for charmonium and bottomonium is calculated. 

For charmonium, the quantum numbers of their lightest hybrid multiplet coincide with ours ($1(s/d)_1$) and the masses are compatible  with ours for the $1^{-+}$ and $2^{-+}$ states within errors (between $150$ MeV and $230$ MeV), but below for the $0^{-+}$ and $1^{--}$ states. For spin zero hybrids, they obtain a $2^{++}$ state $(1(p/f)_2)$ as the second lighter state whereas we have a $1^{++}$ state ($1p_1$). The mass of the $2^{++}$ state is, nevertheless,  compatible with ours within the large errors, but the masses of the $1^{++}$ and $0^{++}$ states are higher. The masses of the spin one hybrids $0^{+-}$ and $1^{+-}$ are compatible, again within large errors.

For bottomonium, they obtain the same hierarchy of multiplets as in charmonium. However, the larger errors make it now compatible with ours, even though the central values are not. The masses of the lightest multiplet are considerably lower than ours, but the ones of the remaining multiplets ($1^{++}$, $0^{+-}$, $1^{+-}$; $2^{++}$; $0^{++}$) are compatible within large errors.

\section{Comparison with experiment}
\label{sec:exp}

In this section, we compare experimental results with ours in the case of maximum mixing. That is with the results displayed on the 6th column of Table \ref{c1--} and on the 4th column of Tables \ref{c0++}-\ref{b2-+}. As mentioned before, the shifts in the spectrum due to mixing are not very important. However, the violations of heavy quark spin 
symmetry induced by the mixing
are crucial to map our results to the XYZ states.
We omit in the analysis the neutral states that have been identified as isospin partners of charged states. 

\subsection{Charm}

\begin{itemize}
\item $X(3823)$ \cite{Olive:2016xmw} is compatible with our $2^{--}$ charmonium $1d$ state ($3792$ MeV).
\item $X(3872)$ \cite{Olive:2016xmw} is compatible with our $1^{++}$ charmonium $2p$ states ($3967$ MeV). Since it sits at the $D^0{\bar D}^{0\ast}$ threshold, it is expected to have a large mixing with those states that we have not taken into account.
\item $X(3915)$ and $X(3940)$ \cite{Olive:2016xmw} are also compatible with our charmonium $2p$ states ($3968$ MeV). Since they are close to the $D_s{\bar D}_s$ threshold ($3936$ MeV), the $0^+$ states may have a large mixing with those states.
\item $Y(4008)$ \cite{Liu:2013dau} is compatible with our $1^{--}$ hybrid $1(s/d)_1$ ($H_1$) state ($4004$ MeV). It mixes with spin one charmonium (see column 7 in table \ref{c1--}
and fig. \ref{1--1sd1}), which may explain the observed spin symmetry violating decays.

\begin{figure}[H]
		\centering 
		\includegraphics[width=0.45\textwidth]{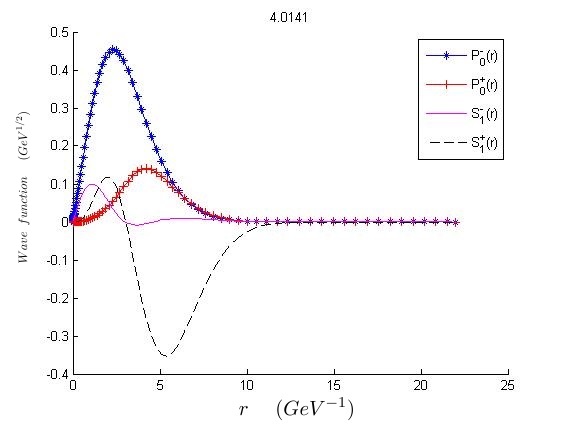}
		\caption{The wave function of the  charmonium $1^{--}$ $1(s/d)_1$ state.}
		\label{1--1sd1}	
	\end{figure}

\item $X(4140)$ \cite{1606.07895} and $X(4160)$ \cite{Olive:2016xmw} are compatible with our $1^{++}$ hybrid $1p_1$ ($H_2$) state ($4146$ MeV). Since the quantum numbers of $X(4160)$ have not been established, it may also correspond to the $1(p/f)_2$ hybrid or to the scalar $3s$ or $2d$ states. The fact that no decays to charmonium of the $1p_1$ state are allowed at leading order is consistent with the fact that no such decays have been observed so far for $X(4160)$, which selects it as our favorite hybrid candidate for that state. 
If so there is no room for the $X(4140)$ ($1^{++}$) in our spectrum. 
These states may be affected by the 
 $D_s^\ast {\bar D}_s$ threshold ($4080$ MeV).

\item $X(4230)$ and $Y(4260)$ \cite{Olive:2016xmw} are compatible with our $1^{--}$ charmonium $2d$ state ($4180$ MeV). It may have a dominant spin zero hybrid component (see table \ref{c1--}), which may help to understand the recent results by the BESSIII collaboration \cite{1611.04669}. Indeed, in \cite{1611.01317} it is claimed that the former $Y(4260)$ peak observed in $\pi^+\pi^- J/\psi$ invariant mass actually consists of two resonances $Y(4220)$ and $Y(4390)$. The parameters of the first resonance are compatible with the ones of $X(4230)$. They are also compatible with the ones of one of the structures observed in $\pi^+\pi^- h_c$ \cite{1610.07044}.  The large hybrid component (see fig. \ref{1--2d}) may explain why it is also observed in this second channel, which would be suppressed by spin symmetry otherwise. It may also be affected by the $D_1 {\bar D}$ threshold ($4290$ MeV).

\begin{figure}[H]
		\centering 
		\includegraphics[width=0.45\textwidth]{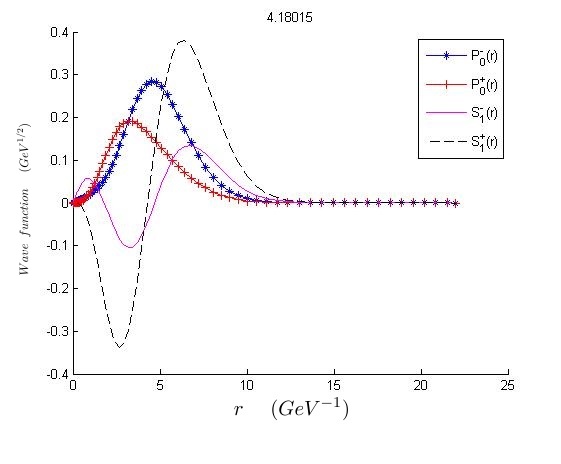}
		\caption{The wave function of the  charmonium $1^{--}$ $2d$ state.}
		\label{1--2d}	
	\end{figure}
	
\item $Y(4274)$ \cite{1606.07895} is compatible with our $1^{++}$ charmonium $3p$ state ($4368$ MeV). 
It may be affected by the $D_s^\ast {\bar D}_s^\ast$ threshold ($4224$ MeV).
\item $X(4350)$ \cite{Olive:2016xmw} is compatible with our spin one $2(s/d)_1$ hybrid states ($4355$ MeV) and charmonium $3p$ states ($4369$ MeV).
\item $Y(4320)$, $Y(4360)$ and $Y(4390)$ \cite{Olive:2016xmw,1611.01317,1610.07044} are compatible with our spin zero $1^{--}$ hybrid $2(s/d)_1$ ($H_1'$) state ($4366$ MeV). Spin symmetry would in principle favor the latter, as it is observed in the $\pi^+\pi^- h_c$ channel rather than in the $\pi^+\pi^- J/\psi$ channel. However, the large mixing with spin one charmonium (see table \ref{c1--} and fig. \ref{1--2sd1}) makes the two first ones also acceptable. The absence of any other state in this region in table \ref{c1--} leaves two of them with no assignment. 
They may be affected by the $D_0^\ast {\bar D}^\ast$ threshold ($4407$ MeV).

\begin{figure}[H]
		\centering 
		\includegraphics[width=0.45\textwidth]{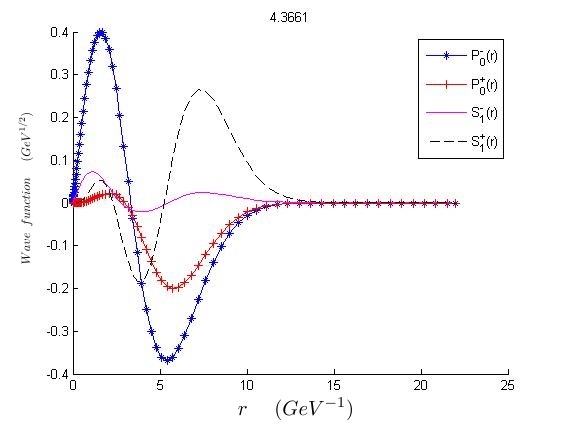}
		\caption{The wave function of the  charmonium $1^{--}$ $2(s/d)_1$ state.}
		\label{1--2sd1}	
	\end{figure}

\item $X(4500)$ \cite{1606.07895} is compatible with our $0^{++}$ hybrid $1p_0$ ($H_3$) state ($4566$ MeV). However, it mixes very little with spin one charmonium (see table \ref{c0++} and fig. \ref{0++1p0}), which does not help to understand the observation in the $J/\psi \phi$ channel.
It may be affected by the $D(2550) {\bar D}^\ast$ threshold ($4557$ MeV).

\begin{figure}[H]
		\centering 
		\includegraphics[width=0.45\textwidth]{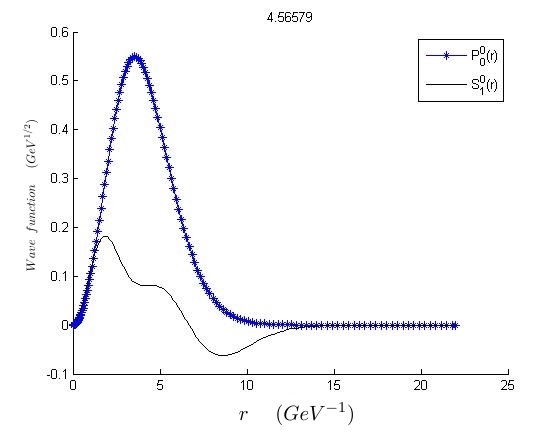}
		\caption{The wave function of the  charmonium $0^{++}$ $1p_0$ state.}
		\label{0++1p0}	
	\end{figure}
	
\item $Y(4630)$ \cite{Olive:2016xmw} is compatible with our $1^{--}$ charmonium $3d$ state ($4559$ MeV). It may be affected by the $D_{s1} {\bar D}_s^\ast$ thresholds ($4572$ MeV and $4648$ MeV).
\item $Y(4660)$ \cite{Olive:2016xmw} is compatible with  our spin zero $1^{--}$ hybrid $3(s/d)_1$ ($H_1''$) state ($4711$ MeV). The mixing with spin one charmonium (see table \ref{c1--}
and fig. \ref{1--3sd1}) may explain the observed decays to vector charmonium. It may be affected by the $D_{s1} {\bar D}_s^\ast$ and $D_{s2}^\ast {\bar D_s}^\ast$ thresholds ($4648$ MeV and $4685$ MeV).

\begin{figure}[H]
		\centering 
		\includegraphics[width=0.45\textwidth]{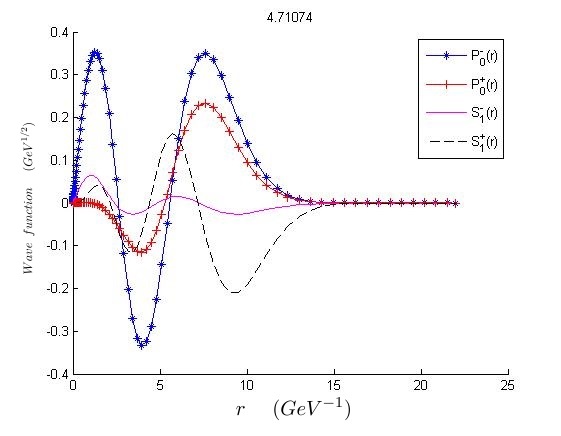}
		\caption{The wave function of the  charmonium $1^{--}$ $3(s/d)_1$ state.}
		\label{1--3sd1}	
	\end{figure}

\item $X(4700)$ \cite{1606.07895} is compatible with our $0^{++}$ charmonium $4p$ state ($4703$ MeV).
\end{itemize}

\begin{widetext}
	
	\begin{figure}

\includegraphics[scale=0.85]{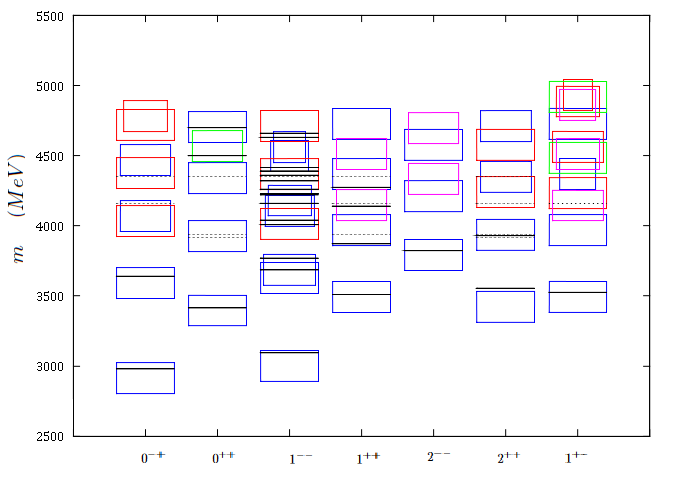}
\caption{Charmonium spectrum including hybrids. The height of the boxes corresponds to the error estimated at the end of Sec. \ref{sp}. Blue boxes correspond to quarkonium, red boxes to $(s/d)_1$ and $(p/f)_2$ hybrids, cyan boxes to $p_1$ and $d_2$ hybrids, and green boxes to $p_0$ hybrids. The black lines are experimental resonances assigned acoording to the discussion in Sec. \ref{sec:exp}. Solid (dashed) lines are resonances with a single (multiple) possible assigment(s). The width of the boxes is chosen arbitrarily in order to facilitate identifications.   }
\label{mix}       
\end{figure}

\end{widetext}

The assignments above can be visualized in Fig. \ref{mix}.

\subsection{Bottom}

\begin{itemize}

\item $\Upsilon (10860)$ \cite{Olive:2016xmw,1501.01137} is compatible with our $1^{--}$ bottomonium $5s$ state ($10881$ MeV). Upon mixing it becomes lighter than the spin zero $2(s/d)_1$ hybrid nearby (see table \ref{b1--} and fig. \ref{1--5s(b)}). Mixing may also explain the large spin symmetry violating decays to $\pi^+\pi^- h_b$ \cite{1508.06562}.

\begin{figure}[H]
		\centering 
		\includegraphics[width=0.45\textwidth]{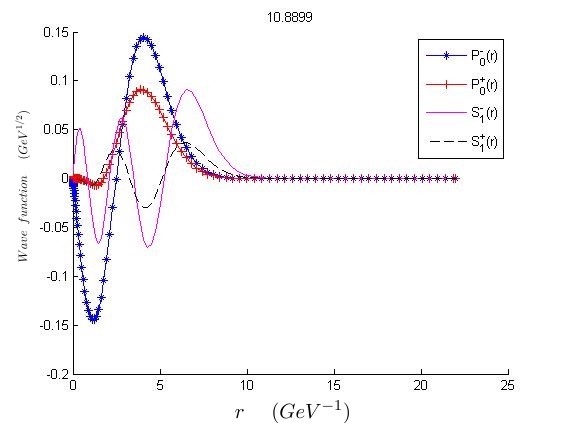}
		\caption{The wave function of the bottomonium $1^{--}$ $5s$ state.}
		\label{1--5s(b)}	
	\end{figure}

\item $Y_b(10890)$ \cite{0808.2445} is compatible with our spin zero $1^{--}$ hybrid $2(s/d)_1$ state ($10890$ MeV). Upon mixing becomes heavier than the $5s$ bottomonium nearby (see table \ref{b1--} and fig. \ref{1--2sd1(b)}).

\begin{figure}[H]
		\centering 
		\includegraphics[width=0.45\textwidth]{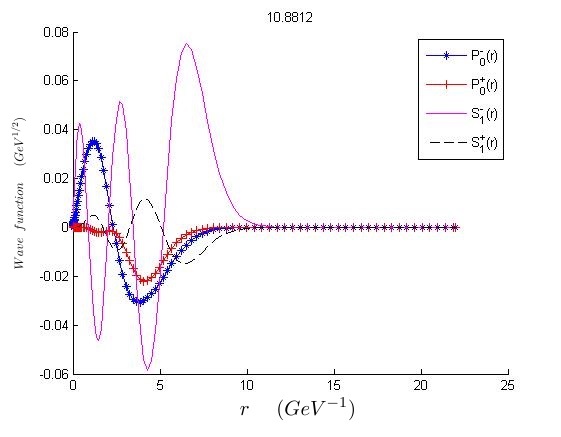}
		\caption{The wave function of the bottomonium $1^{--}$ $2(s/d)_1$ state.}
		\label{1--2sd1(b)}	
	\end{figure}

\item $\Upsilon (11020)$ \cite{Olive:2016xmw,1501.01137} is about $1\sigma$ heavier than our $1^{--}$ bottomonium $4d$ state ($10942$ MeV). It may be affected by the $B_1 {\bar B}$ threshold ($11000$ MeV). 

\end{itemize}

\section{Discussion}
\label{discussion}

We have compared our results to other QCD based approaches in section \ref{sec:th}. We find good agreement with Born-Oppenheimer approaches that have appeared recently in the literature \cite{Braaten:2014qka,Berwein:2015vca}, as expected. However the agreement with QCD sum rules and lattice QCD calculations is marginal. The lattice calculations in anisotropic lattices and unphysical quark masses tend to give a heavier hybrid spectrum, both in relativistic implementations of heavy quarks \cite{Cheung:2016bym,hep-lat/0111049} as well as in lattice NRQCD \cite{Juge:1999ie}. Nevertheless in \cite{1603.06467}, a lattice calculation in which both the continuum and the chiral extrapolations are carried out, the three states found that are not identified with known quarkonia fit well in our spectrum. In particular, the $1^{--}$ state is compatible with the one in our lightest hybrid multiplet. 

It is remarkable that the gross features of the experimental charmonium and bottomonium spectrum, including isopin zero XYZ states, can be understood from our results. The main improvement with respect to previous works is that in addition to the Cornell potential for the quarkonium sector and the Born-Oppenheimer potentials for the hybrid sector, we include the leading mixing term between those sectors. The mixing term implies that the actual physical states are a superposition of spin zero (one) hybrids and spin one (zero) quarkonium. This facilitates the identification of certain Y states as hybrids, since otherwise the apparent spin symmetry violating decays were difficult to understand \cite{Berwein:2015vca}. We would like to emphasize that the mixing term we use is essentially derived from NRQCD, and hence from QCD. Its short and long distance behavior are obtained in a model independent way. The model dependence comes in through the interpolation we use. We have chosen the sign combination and a value of the free parameter such that a large mixing is favored. It would be very important to have a lattice evaluation of the mixing potential to validate these choices (or otherwise). We have produced formulas (\ref{lattice}) that can be easily implemented on the lattice (see for instance \cite{Koma:2006si,Koma:2006fw}).

There appear to be too many known isospin zero $1^{--}$ charmonium resonances to fit our spectrum in table \ref{c1--} (see also fig. \ref{mix}). If we assign the $Y(4008)$ to the $1(s/d)_1$ state, then $\psi (4040)$ and 
$\psi (4160)$ naturally fall into the $3s$ and $2d$ states respectively. However, the $X(4230)/Y(4220)$ are also candidates for the $2d$ state. A possible way out would be to disregard
$Y(4008)$, as it is a very wide resonance that has only been observed by Belle. Then $\psi (4040)$ would be assigned to the $1(s/d)_1$ state, $\psi (4160)$ to the $3s$ state and $X(4230)/Y(4220)$ to the $2d$. 
The fact that the $1(s/d)_1$ state has about a $30\%$ quarkonium component according to the 7th column of table \ref{c1--} (see also fig. \ref{1--1sd1}) may explain why it has been labeled as $\psi (4040)$. For the next state, $2(s/d)_1$,
 there are three competing resonances $Y(4320)$, $Y(4360)$ and $Y(4390)$. This makes us suspect that they could well correspond to the same state. Indeed, the decay widths of $Y(4320)$ and $Y(4360)$ are 
compatible and the one of $Y(4390)$ is less than $1\sigma$ away. Concerning the masses, $Y(4320)$ and $Y(4360)$ are less than  $1\sigma$ away, but $Y(4390)$ is more than $5\sigma$ away, which casts some 
doubts on the suggested identification. Leaving this puzzle aside, there would only be one state to be discovered below the $Y(4660)$, the $3d$ around $4560$ MeV.

If we assume for the $2(s/d)_1$ states the mixing displayed in column 7 of table \ref{c1--} and the decay width in table \ref{decayW} for the hybrid component, we obtain
\be
\Gamma \left(Y(4320/4360/4390) \to h_c + {\rm l.h.}\right)= 14 (12)\, {\rm MeV} \,,
\ee  
where l.h. stands for light hadrons. Analogously, for the $X(4230)/Y(4220)$ state we have,
\be
\Gamma \left(X(4230)/Y(4220) \to h_c + {\rm l. h.}\right)= 17 (15)\, {\rm MeV} \,.
\ee  

Concerning the $1^{--}$ bottomonium resonances, all of them fit in our spectrum in table \ref{b1--}. In addition, there should be three states still to be discovered below the $\Upsilon (10860)$, 
the $2d$, $1(s/d)_1$ and $3d$ around $10440$ MeV, $10690$ MeV and $10710$ MeV respectively.

If we take the mixing in column 5 of table \ref{b1--} and the decay width in table \ref{decayW} for the hybrid component, 
we can also estimate the following decay widths for bottomonium,  
\bea
\Gamma \left(\Upsilon(10860) \to h_b + {\rm l.h.}\right) &=& 3 (1)\, {\rm MeV} \nn\\
\Gamma \left(Y_b(10890) \to h_b + {\rm l.h.}\right) &=& 13 (6)\, {\rm MeV}\, .
\eea

According to our identifications in section \ref{sec:exp}, we can infer the quantum numbers of some XYZ states. 
\begin{itemize}
\item $X(3915)$ should be the $\chi_{c0}'$ ($0^{++}$).
\item $X(3940)$ should be the $h_{c}'$ ($1^{+-}$).
\end{itemize}

It is important to keep in mind that there are further $1/m_Q$ corrections to the hybrid spectrum beyond those that induce mixing between hybrids and quarkonia we have focused on. 
In particular, the fine and hyperfine splittings of hybrids may appear at ${\cal O} (1/m_Q)$  rather than at ${\cal O} (1/m_Q^2)$, as those of quarkonium. Indeed, the following terms are compatible with the symmetries of (\ref{symH}),
\bea
&&i\epsilon^{ijk}V^S(r) tr\left(H^{i\dagger}\left[\sigma^k , H^j\right]\right) \,, \\ 
&&i\epsilon^{ijk}V^L(r) tr\left(H^{i\dagger} L^k H^j\right) \,,
\eea
($L^k$ is the angular momentum operator) and may appear at ${\cal O} (1/m_Q)$ in the matching to NRQCD.

Before closing, let us briefly discuss the important question on how the lattice potentials we use (fig. \ref{fig2})
may change in the case $n_f=3$ (three light quarks). 
We know that $\Sigma_g^+$ does not change
much and this is also so for $\Pi_u$ \cite{Bali:2000vr}, at least up to moderately large distances.
 Nothing is known about $\Sigma_u^-$, but there is no reason 
to expect a different behavior. Two major qualitative features arise though. The first one is 
the appearance 
of heavy-light meson pairs, which amount to roughly horizontal lines at the threshold energies in fig. \ref{fig2}. 
These states interact with the remaining potentials already at leading order, and may in principle 
produce important distortions with respect to the $n_f=0$ case.
In practice, we only know how they cross talk to the $\Sigma_g^+$ state, and turn out to produce a tiny
 disturbance to the spectrum, apart from avoiding level crossing \cite{Bali:2005fu}. Hence we 
expect the effects of
 $n_f\not=0$ to be important only when our states are very close to some heavy-light meson pair 
threshold. This is the reason why we quoted the location of nearby thresholds when identifying our hybrid candidates with XYZ states in section \ref{sec:exp}.
The second one is the appearance of light quark excitations, in addition to the gluon ones, in 
the 
static spectrum of fig. \ref{fig2}. They may have different quantum numbers, for instance non zero isospin (in this
case they may be relevant to the experimentally discovered charged $Z$ states). We do not know 
anything about those and, as pointed out in \cite{Brambilla:2008zz} 
and more recently emphasized in \cite{Braaten:2013boa,Braaten:2014qka}, it would be extremely 
important to have lattice QCD evaluations of the static energies of light quark excitations. We suspect that light quark excitations with the same quantum numbers as the gluonic ones will only provide small modifications 
to the hybrid potentials, since they correspond to higher dimensional operators. 
In this respect, it is significant that tetraquark models have also difficulties to encompass the $X(4140)$ in their spectrum together with $X(4237)$, $X(4500)$ and $X(4700)$ \cite{Esposito:2016noz}. In fact, the $X(4140)$ structure may be due to a threshold enhancement according to some authors \cite{vanBeveren:2009dc,Liu:2016onn,Ortega:2016hde}.
This means that tetraquarks with the same quantum numbers as hybrids will in general be hidden in the spectrum of the latter.

\section{Conclusions}
We have calculated the charmonium and bottomonium hybrid spectrum in a QCD based approach, including for the first time the mixing with standard charmonium and bottomonium states. The latter leads to enhanced spin symmetry violations, which are instrumental to identify a number of XYZ states as hybrid states. Most of the isospin zero XYZ states fit well in our spectrum, either as hybrids or as standard quarkonium states. We have also estimated several decay widths.
\label{sec:concl}

\begin{acknowledgments}
J.S. thanks Josep Taron for collaboration in the early stages of this work, Jaume Tarrús for discussions and Antonio Pineda for providing the data used in ref. \cite{Bali:2003jq}. We have been supported by the  Spanish Excellence Network on
Hadronic Physics FIS2014-57026-REDT.  J.S. also acknowledges support from
the 2014-SGR-104 grant (Catalonia), the FPA2013-4657, FPA2013-43425-P, FPA2016-81114-P and FPA2016-76005-C2-1-P projects (Spain).

\end{acknowledgments}

\begin{appendices}
	\appendix

	\section{Quarkonium}
	\label{quarkonium}
	
	Conventional quarkonium, namely $Q\bar{Q}$ in a color singlet state, can be described by the Schrödinger equation using the ground state potential $V_{\Sigma_g^+}(r)$. 
	
	\begin{equation}
	h=-\frac{\boldsymbol{\nabla^2}}{m_Q}+V_{\Sigma_g^+}(r)\,.
	\end{equation}
We approximate $V_{\Sigma_g^+}(r)$ by the Cornell potential,	 
	\begin{equation}
\label{V_g}
V_{\Sigma_{g}^{+}}(r)\approx -\frac{k_g}{r}+\sigma_g r+E_g^{Q\bar{Q}} \, ,
\end{equation} 
where we take,
\be
k_g=0.489 \, ,\hspace{1cm} \sigma_g=0.187 GeV^2 \, ,
\ee 
which describes lattice data well, see fig. \ref{F1}. $E_g^{Q\bar{Q}}$ will be tuned independently for charmonium and bottomonium. 
We write the wave-function as $S({\bf r})=
	\frac{R_L(r)}{r}Y_{LM}(\theta,\phi)$, which leads to the reduced equation:
	
	\begin{equation}
	\label{SRq=0}
	\left(-\frac{1}{m_Q}\frac{\partial^2}{\partial r^2}+\frac{L(L+1)}{m_Q r^2}+V_{\Sigma_g^+}(r)\right)R_L(r)=E R_L(r)\,.
	\end{equation}
	The different eigenvalues of this equation correspond to the energy levels of heavy quarkonium, many of which have been experimentally confirmed for charmonium and bottomonium \cite{Olive:2016xmw}.
 We fix $E_g^{Q\bar{Q}}$ by making the charmonium and bottomonium spectrum to best agree with the respective experimental spin averages. We obtain, 
	\be	
	E_g^{c\bar{c}}=
	-0.242GeV \hspace{1cm} E_g^{b\bar{b}}=
	-0.228GeV \, .
	\ee
	The table \ref{qqresults}  shows the results in terms of $M_{Q\bar{Q}}=2m_Q + E$ for $Q=c\,, b$ of eq.(\ref{SRq=0}) for the lower $nL$ 
	energy states. It also shows the expectation value of the momentum, the inverse radius, the expected size of the higher order corrections and our error estimate. $V^{(1)}$, $V^{(2)}_{vd}$ (velocity dependent) and  $V^{(2)}_{vi}$ (velocity independent) depend on $\lQ$ and $r$. We take $\lQ = 400$ MeV and estimate them as follows. If $\lQ > 1/\left<r\right>$ we take them as $\lQ^2$, $\lQ\left< p\right>^2$ and $\lQ^3$ respectively. If $\lQ < 1/\left<r\right>$ we take them according to the weak coupling scaling
	$\als^2/\left<r\right>^2$, $\als  \left< p\right>^2/\left<r\right>$ and $\als / \left<r\right>^3$ respectively, where $\als$ is the one loop running coupling constant evaluated at the scale $1/\left<r\right>$. The total error is obtained by summing in quadrature these estimates and the relativistic correction to the kinetic energy displayed in the eighth column. We observe that the errors for charmonium are rather large, and are dominated by the velocity dependent potential. We also display the experimental results in the last column.

	\begin{table}[h]
	\centering
	\begin{tabular}{|c|c|c|c|c|c|c|c|c|c|}    
		\hline                                                                     
		$nL$  &  $M_{Q\bar{Q}}$      & $<p>$    &  $\frac{1}{<r>}$ & $\frac{V^{(1)}}{m_Q}$ &  $\frac{V^{(2)}_{vd}}{m_Q^2}$ & $\frac{V^{(2)}_{vi}}{m_Q^2}$ & $\frac{p^4}{8m_Q^3}$   & $\Delta M_{Q\bar{Q}}$  & $E_{exp}$ \\
		\hline \hline
$1s$ & 3068 & 738 & 518 & 54 & 71 & 35 & 12 & 96 & 3068 \\
$2s$ & 3678 & 836 & 259 & 109 & 129 & 30 & 19 & 173 & 3674 \\ 
$3s$ & 4130 & 935 & 186 & 109 & 162 & 30 & 30 & 199 & 4039 \\ 
$4s$ & 4517 & 1019 & 149 & 109 & 192 & 30 & 42 & 227 & 4421 \\ 
$5s$ & 4865 & 1097 & 127 & 109 & 223 & 30 & 57 & 256 & ? \\ 
$1p$ & 3494 & 753 & 317 & 109 & 105 & 30 & 13 & 155 & 3525 \\ 
$2p$ & 3968 & 871 & 209 & 109 & 140 & 30 & 23 & 182 & 3927 \\ 
$3p$ & 4369 & 966 & 162 & 109 & 173 & 30 & 34 & 209 & ? \\ 
$4p$ & 4726 & 1048 & 135 & 109 & 203 & 30 & 47 & 237 & ? \\ 
$5p$ & 5055 & 1136 & 119 & 109 & 239 & 30 & 66 & 272 & {?} \\ \hline
$1s$ & 9442 & 1546 & 1028 & 29 & 37 & 17 & 6 & 50 & 9445 \\ 
$2s$ & 10009 & 1408 & 432 & 14 & 22 & 2 & 4 & 26 & 10017 \\ 
$3s$ & 10356 & 1494 & 295 & 33 & 38 & 3 & 5 & 50 & 10355 \\ 
$4s$ & 10638 & 1594 & 232 & 33 & 43 & 3 & 7 & 54 & 10579 \\ 
$5s$ & 10885 & 1692 & 195 & 33 & 48 & 3 & 9 & 59 & 10876 \\ 
$1p$ & 9908 & 1268 & 531 & 17 & 19 & 3 & 3 & 26 & 9900 \\ 
$2p$ & 10265 & 1386 & 332 & 33 & 32 & 3 & 4 & 46 & 10260 \\ 
$3p$ & 10553 & 1504 & 252 & 33 & 38 & 3 & 5 & 51 & {?} \\ 
$4p$ & 10806 & 1612 & 207 & 33 & 44 & 3 & 7 & 55 & {?} \\ 
$5p$ & 11035 & 1727 & 180 & 33 & 50 & 3 & 10 & 61 & {?} \\ \hline
	\end{tabular}
	\caption{Masses, average momentum, inverse radius, expected sizes of higher order contributions ($1/m_Q$ potential, $1/m_Q^2$ velocity dependent potential, $1/m_Q^2$ velocity independent potentials, $1/m_Q^3$ kinetic energy) and estimated error (in MeV) for charmonium (upper) and bottomonium (lower). The error is estimated by summing in quadrature the expected sizes of the higher order contributions (see the text for details on the latter). We have taken $m_c=1.47$ GeV and  $m_b=4.88$ GeV. The experimental numbers are displayed in the last column.}
	\label{qqresults}
\end{table}

	\section{Extraction of $g\Lambda'$ and $g\Lambda'''$ from lattice data}
	\label{fits}
	
	$g\Lambda'$ and $g\Lambda'''$ also appear in the $1/m_Q^2$ quarkonium potentials \cite{PerezNadal:2008vm,Brambilla:2014eaa}. Following the notation of ref. \cite{Koma:2009ws}, we have that the 
long distance behavior of the spin-orbit,
	tensor, and spin-spin potentials reads,
	\bea
	&&V_2'(r)=2rV_{{\bf L}_2{\bf S}_1}^{(1,1)}=-\frac{2c_F g^2\Lambda^2\Lambda'}{\kappa r}=-\frac{2c_F g\Lambda'}{r}\nn\\
	&&V_3 (r)= 12 V_{{\bf S}_{12}}^{(1,1)}=\frac{2\pi^3 c_F^2 g^2 {\Lambda'''}^2}{15\kappa^2 r^5}\\
	&&V_4 (r)= 3 V_{{\bf S}^2}^{(1,1)}=\frac{\pi^3 c_F^2 g^2 {\Lambda'''}^2}{30\kappa^2 r^5} \quad .\nn
	\eea
	We shall take the tree level value for $c_F$, $c_F=1$.
	
\begin{figure}[H]
	\centering 
	\includegraphics[width=0.45\textwidth]{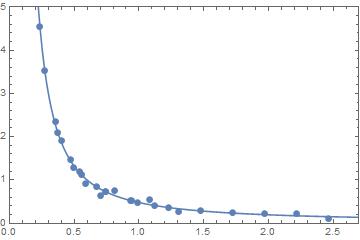}
	\caption{$V_2'(r)$ in units of $r_o^{-2}$ against $r$ units of $r_o$, $r_o\sim0.5$ fm.}
	\label{V2}	
\end{figure}

	For the spin-orbit potential, a simple interpolation of the expected long and short distance behavior, namely
	\be
	V_2'(r)= \frac{A}{r^2}+\frac{B}{r} \,,
	\ee
	already produces a good fit to data ($R^2=0.998$, see fig. \ref{V2}). We obtain $A=0.181$ and $B=0.295$ in units of $r_0$, which translates to $\vert g\Lambda'\vert=0.059$ GeV.
If we restrict ourselves to the longer distance points (from  seven
to three) and fit the expected long distance behavior only, we obtain worse fits ($R^2\lesssim 0.977$) with numbers about a $40\%$ higher, which may serve to estimate the error.

\begin{figure}[H]
	\centering 
	\includegraphics[width=0.45\textwidth]{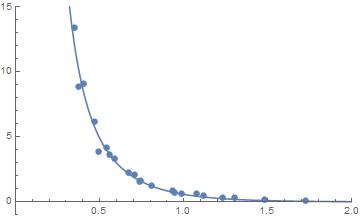}
	\caption{$V_3(r)$ in units of $r_o^{-3}$ against $r$ in units of $r_o$, $r_o\sim0.5$ fm.}
	\label{V3}	
\end{figure}

        For the tensor potential, the following interpolation, which also has the right short and long distance behavior, produces a good fit to data ($R^2=0.996$, see fig. \ref{V3}),
\be
V_3(r)=\frac{C+Dr}{r^3+r^6}\, .
\ee
        We obtain $C=0.191$ and $D=1.00$ in units of $r_0$, which translates to $\vert g\Lambda'''\vert= 0.230$ GeV. We have checked that if we restrict ourselves to the longer distance points (from  seven
to three) and fit the expected long distance behavior only, we obtain numbers compatible with the latter within a $35\%$ error.
$\vert g\Lambda'''\vert$ may also be obtained from the long distance behavior of the spin-spin potential. However,
we have not been able to find a good fit to the data of ref. \cite{Koma:2009ws}, neither using simple interpolations between the expected short and long distance behavior nor to the expected long distance behavior for the longer distance points 
(from nine to three).

	\section{Tensor Spherical Harmonics}
	\label{tensor}
	
	We follow the notation of ref. \cite{pg}. We define
	\be
	Y^{ijLJ}_{\cal J M}=\sum_{\nu=0,\pm 1} C(J1{\cal J};{\cal M}-\nu\, \nu) Y^{iL}_{J\,{\cal M}-\nu}\chi^j_\nu \,,
	\ee
	where $Y^{iL}_{J\,M}$ are the vector spherical harmonics,
	\be
	Y^{iL}_{J\,{M}}=\sum_{\mu=0,\pm 1} C(L1J;M-\mu\, \mu) Y^{M-\mu}_L \chi^i_\mu \,,
	\ee
	where $ Y^{M}_L$ are the usual spherical harmonics and 
	\be
	{\boldsymbol{\chi}}_{\pm 1}=\mp\frac{1}{\sqrt{2}}
	\begin{pmatrix} 
	1 \\ \pm i\\ 0
	\end{pmatrix}
	\quad ,\quad 
	{\boldsymbol{\chi}}_{0}=
	\begin{pmatrix}
	0\\0\\ 1 
	\end{pmatrix} \, .
	\ee
	$C(J_1 J_2 J;M_1 M_2)$ are the Clebsch-Gordan coefficients.
	
	\section{Spectrum}
	\label{mixspectrum}
	
	We display in this appendix the tables for the full charmonium and bottomonium spectrum up to ${\cal J}=2$, which includes hybrids and quarkonia states, except
	for the charmonium $1^{--}$ case that is displayed in table \ref{c1--}.
	
	\begin{table}[htbp]
	\centering
	
	\begin{tabular}{|c|c|c|c|c|}
	\hline
	$NL_J$ & $\lambda=0$ & \%   & $\lambda=0.6$  &\%    \\
	\hline 
	$1p$ & 3.494 &  0&3.396 &  7 \\ 
		$2p$ & 3.968 & 0 & 3.925 & 1 \\ 
		$3p$ & 4.369 &  0& 4.338 & 0 \\
	$1p_0$ & 4.486 & 100& 4.566 & 98 \\ 
	$4p$ & 4.727 & 0 & 4.703 & 9 \\ 
	
		$2p_0$ & 4.920 & 100 & 4.965 &  94\\ 
		$5p$ & 5.055 & 0 & 5.034& 1\\ \hline
	\end{tabular}
	\caption{Spectrum of charmonium ($S=1$) and hybrids ($S=0$): $0^{++}$ states. Masses are in GeV. The \% columns show the fraction of the hybrid components for the mass states in the previous column. The mixing potentials are fixed to $V_S^\Pi [+-]$ and $V_S^\Sigma [++]$. $m_c=1.47$ GeV. }
	
	\label{c0++}
\end{table}

\begin{table}[htbp]
	\centering
\begin{tabular}{|c|c|c|c|c|}
	
	\hline
	$NL_J$ & $\lambda=0$ & \%   & $\lambda=0.6$  &\%   \\
	\hline 
	$1p$ & 3.494 &  0&3.492 &  0 \\ 
		$2p$ & 3.968 & 0 & 3.967 & 0 \\ 
	
		$1p_1$ & 4.145 & 100 & 4.146 & 100 \\ 
		
		$3p$ & 4.369 &  0& 4.368 & 0 \\
		$2p_1$ & 4.511 & 100 & 4.512 & 100 \\ 
		$4p$ & 4.727 & 0 & 4.726 & 0 \\ 
		
		$3p_1$ & 4.863 & 100 & 4.863 & 99 \\ 
 
		$5p$ & 5.055 & 0 & 5.055& 1\\ \hline
	\end{tabular}
	\caption{Same as in table \ref{c0++} for
	$1^{++}$ states.}
	
	\label{c1++}
\end{table}

\begin{table}[htbp]
	\centering
\begin{tabular}{|c|c|c|c|c|}
	
	\hline
	$NL_J$ & $\lambda=0$ & \%   & $\lambda=0.6$  &\%   \\
	\hline 
	$1p$ & 3.494 &  0&3.424 &  5 \\ 
		$2p$ & 3.968 & 0 & 3.937 & 7 \\ 
		$1f$ & 4.047 & 0 & 3.981 & 11 \\
		$1(p/f)_2$ & 4.231 & 100 & 4.240 & 81 \\
		$3p$ & 4.369 &  0& 4.350 & 0 \\
    	$2f$ & 4.428 &  0& 4.391 & 77 \\
		$2(p/f)_2$ & 4.563 & 100 & 4.579 & 53 \\
		$4p$ & 4.727 & 0 & 4.709 & 3 \\ 
		
		$3f$ & 4.775 & 0 & 4.752 & 11 \\
		$3(p/f)_2$ & 4.886 & 100 & 4.909 & 78 \\ 
		$4(p/f)_2$ & 4923 & 100 & 4.952 & 94\\
		$5p$ & 5.055 & 0 & 5.040& 4\\ 
		
		\hline
	\end{tabular}
	\caption{Same as in table \ref{c0++} for
	$2^{++}$ states.}
	\label{c2++}
\end{table}

\begin{table}[htbp]
	\centering
	
	\begin{tabular}{|c|c|c|c|c|}
		\hline
		$NL_J$ & $\lambda=0$ & \%   & $\lambda=0.6$  &\%   \\
		 \hline 
		$1d$ & 3.793 & 0 & 3.792 & 0 \\	  
	$2d$ & 4.210 & 0 & 4.209 & 1 \\
	$1d_2$ & 4.334 & 100 & 4.335 & 100 \\	
		$3d$ & 4.579 & 0 & 4.578 & 0 \\	
	$2d_2$ & 4.693 & 100 & 4.694 & 99\\	
	$4d$ & 4.916 & 0  & 4.915 & 0 \\ 
$3d_2$ & 5.036 &	100 & 5.037 &  100\\
		 \hline
	\end{tabular}
	\caption{Same as in table \ref{c0++} for 
	$2^{--}$ states.} 
	\label{c2--}
\end{table}

\begin{table}[htbp]
	\centering
	
	\begin{tabular}{|c|c|c|c|c|}
		\hline
		$NL_J$ & $\lambda=0$ & \%   & $\lambda=0.6$  &\%   \\
		 \hline 
		$1s$ & 3.068 & 0 & 2.913 & 7 \\
		$2s$ & 3.678 & 0 & 3.591 & 8 \\ 
		$1(s/d)_1$ &  4.011 & 100 & 4.033  & 99 \\
		$3s$ & 4.131 & 0 & 4.069 &1 \\
		$2(s/d)_1$ & 4.355 & 100 & 4.375 & 92 \\
		$4s$ & 4.512 & 0 & 4.468 & 7 \\
		$3(s/d)_1$ & 4.692 & 100 & 4.719 & 99 \\
		$4(s/d)_1$ & 4.718 & 100 & 4.781 & 96 \\ 
	$5s$& 4.865 & 0	 & 4.823 & 0 \\
		$5(s/d)_1$ & 5.043 & 100 & 5.055 & 96 \\
		 \hline
	\end{tabular}
	\caption{Spectrum of charmonium ($S=0$) and charmonium hybrids ($S=1$): $0^{-+}$ states. Masses are in GeV. The \% columns show the fraction of the hybrid components for the mass states in the previous column. The mixing potentials are fixed to $V_S^\Pi [+-]$ and $V_S^\Sigma [++]$.	$m_c=1.47$ GeV.} 
	
	\label{c0-+}
\end{table}

\begin{table}[htbp]
	\centering
	
	\begin{tabular}{|c|c|c|c|c|}
	
	\hline
	$NL_J$ & $\lambda=0$ & \%   & $\lambda=0.6$  &\%    \\
	\hline 
	$1p$ & 3.494 &  0&3.333 &  9 \\ 
		$2p$ & 3.968 & 0 & 3.901 & 2 \\ 
		$1p_1$ & 4145 & 100 & 4.146 & 100\\
		$1(p/f)_2$ & 4231 & 100 & 4.242 & 99\\
		$3p$ & 4.369 &  0 & 4.320 & 1 \\
	$1p_0$ & 4.486 & 100 & 4.511 & 98\\ 
	$2p_1$ & 4.511 & 100 & 4.526 & 100 \\
	$2(p/f)_2$ & 4563 & 100 & 4.590 & 95 \\
	$4p$ & 4.727 & 0 & 4.686 & 8 \\ 
	$3p_1$ & 4863 & 100 & 4.863 & 100 \\
	$3(p/f)_2$ & 4886 & 100 & 4.901 & 99 \\ 
		$2p_0$ & 4920 & 100 & 4.936 &  95\\ 
		$4(p/f)_2$ & 4923 & 100 & 4.959 & 100\\
		$5p$ & 5.055 & 0 & 5.020& 7\\ \hline
	\end{tabular}
	\caption{Same as in table \ref{c0-+} for
	$1^{+-}$ states. }
	\label{c1+-}
\end{table}

\begin{table}[htbp]
	\centering
	
	\begin{tabular}{|c|c|c|c|c|}
		\hline
		$NL_J$ & $\lambda=0$ & \%   & $\lambda=0.6$  &\%   \\
		 \hline 
		$1d$ & 3.793 & 0 & 3.721 & 6 \\
		$1(s/d)_1$ &  4.011 & 100 & 4.014  & 75 \\
	$2d$ & 4.210 & 0 & 4.199 & 80 \\
	$1d_2$ & 4334 & 100 & 4.335 & 100 \\
		$2(s/d)_1$ & 4.355 & 100 & 4.353 & 73 \\ 
		$1(d/g)_3$ & 4.435 & 100 & 4.443 & 100 \\
		$3d$ & 4.579 & 0 & 4.571 & 11 \\
		$3(s/d)_1$ & 4.692 & 100 & 4.690 & 97 \\
	$2d_2$ & 4.693 & 100 & 4.694 & 98\\
		$4(s/d)_1$ & 4.718 & 100 & 4.713 & 96 \\ 
		$2(d/g)_3$ & 4.763 & 100 & 4.774 &  90\\
	$4d$ & 4.916 & 0  & 4.911 & 27 \\ 
$3d_2$ & 5.036 &	100 & 5.037 &  95\\ 
		$5(s/d)_1$ & 5.043 & 100 & 5.084 & 98 \\ 
		 \hline
	\end{tabular}
	\caption{Same as in table \ref{c0-+} for 
	$2^{-+}$ states.} 
	\label{c2-+}
\end{table}

\begin{table}[htbp]
	\centering
	\begin{tabular}{|c|c|c|c|c|}
		\hline
		$NL_J$ & $\lambda=0$ & \%  & $\lambda=0.6$ &\% \\ \hline 
		$1s$       & 9.442   & 0   & 9.441  & 0 \\   
		$2s$       & 10.009  & 0   & 10.000  & 2 \\
		$1d$       & 10.155  & 0   & 10.133  & 2 \\ 
		$3s$       & 10.356  & 0   & 10.352  & 0 \\
		$2d$       & 10.454  & 0   & 10.440  & 2 \\  
		$4s$       & 10.638  & 0   & 10.635  & 1 \\ 
		$1(s/d)_1$ & 10.690  & 100 & 10.688  & 79 \\ 
		$3d$       & 10.712  & 0   & 10.713  & 56 \\
		$2(s/d)_1$ & 10.885  & 100 & 10.881  & 17  \\
		$5s$       & 10.886  & 0   & 10.890  & 75 \\
		$4d$       & 10.947  & 0   & 10.942  & 11  \\ 
		$3(s/d)_1$ & 11.084  & 100 & 11.086  & 98  \\ \hline
	\end{tabular}
	\caption{Spectrum of bottomonium ($S=1$) and hybrids ($S=0$): $1^{--}$ states. Masses are in GeV. The \% columns show the fraction of the hybrid components for the mass states in the previous column. The mixing potentials are fixed to $V_S^\Pi [+-]$ and $V_S^\Sigma [++]$. $m_b=4.88$ GeV.}
	
	\label{b1--}
\end{table}

\begin{table}[htbp]
	\centering
	\begin{tabular}{|c|c|c|c|c|}
		\hline
		$NL_J$ & $\lambda=0$ & \%   & $\lambda=0.6$  &\%    \\
		\hline 
		$1p$       & 9.908  &  0  & 9.907   &  0 \\ 
		$2p$       & 10.265 & 0   & 10.264  & 0 \\ 
		$3p$       & 10.553 & 0   & 10.553  & 0 \\ 
		$4p$       & 10.806 & 0 & 10.805    &  0\\ 
		$1p_0$     & 11.011 & 100 & 11.013  & 99\\ \hline
	\end{tabular}
	\caption{Same as in table \ref{b1--} for $0^{++}$ states.}
	\label{b0++}
\end{table}

\begin{table}[htbp]
	\centering
	\begin{tabular}{|c|c|c|c|c|}
		\hline
		$NL_J$ & $\lambda=0$ & \%   & $\lambda=0.6$  &\%    \\
		\hline 
		$1p$       & 9.908   &  0  & 9.908   &  0 \\ 
		$2p$       & 10.265  & 0   & 10.265  & 0 \\ 
		$3p$       & 10.553  & 0   & 10.553  & 0 \\ 
		$1p_1$     & 10.761  & 100 & 10.761  &  99\\ 
		$1p$       & 10.806  & 0   & 10.806  & 0\\
		$2p_1$     & 10.970  & 100 & 10.970  & 99 \\  
		$5p$       & 11.034  & 0   & 11.035  & 0\\ \hline
	\end{tabular}
	\caption{Same as in table \ref{b1--} for $1^{++}$ states.}
	\label{b1++}
\end{table}

\begin{table}[htbp]
	\centering
	\begin{tabular}{|c|c|c|c|c|}
		\hline
		$NL_J$ & $\lambda=0$ & \%   & $\lambda=0.6$  &\%    \\
		\hline 
		$1p$       & 9.908  &  0  & 9.898   &  1 \\ 
		$2p$       & 10.265 & 0   & 10.258  & 1 \\ 
		$1f$       & 10.348 & 0   & 10.331  & 2 \\ 
		$3p$       & 10.553 & 0   & 10.549  &  0\\ 
		$2f$       & 10.615 & 0   & 10.603  & 5\\
		$4p$       & 10.806 & 0   & 10.801  & 13 \\  
		$1(p/f)_2$ & 10.819 & 100 & 10.820  &  91\\ 
		$3f$       & 10.855 & 0   & 10.851  & 32\\
		$2(p/f)_2$ & 11.005 & 100 & 11.009  & 80\\ \hline
	\end{tabular}
	\caption{Same as in table \ref{b1--} for $2^{++}$ states.}
	\label{b2++}
\end{table}

\begin{table}[htbp]
	\centering
	\begin{tabular}{|c|c|c|c|c|}
		\hline
		$NL_J$ & $\lambda=0$ & \%   & $\lambda=0.6$  &\%    \\
		\hline 
		$1d$       & 10.155  &  0  & 10.155  &  0 \\ 
		$2d$       & 10.453  & 0   & 10.454  & 0 \\ 
		$3d$       & 10.712  & 0   & 10.713  & 0 \\ 
		$1d_2$     & 10.870  & 100 & 10.870  &  100\\ 
		$4d$       & 10.947  & 0   & 10.947  & 0\\
		$2d_2$     & 11.074  & 100 & 10.074  & 100 \\  \hline
	\end{tabular}
	\caption{Same as in table \ref{b1--} for $2^{--}$ states.}
	\label{b2--}
\end{table}

\begin{table}[htbp]
	\centering
	\begin{tabular}{|c|c|c|c|c|}
		\hline
		$NL_J$ & $\lambda=0$ & \%   & $\lambda=0.6$  &\%    \\
		\hline 
		$1s$       & 9.442   &  0  & 9.427   &  1 \\ 
		$2s$       & 10.009  & 0   & 9.987   & 3  \\ 
		$3s$       & 10.356  & 0   & 10.343  & 1\\ 
		$4s$       & 10.638  & 0   & 10.629  &  3 \\ 
		$1(s/d)_1$ & 10.690  & 100 & 10.693  & 99 \\
		$5s$       & 10.886  & 0   & 10.877  & 16\\ 
		$2(s/d)_1$ & 10.885  & 100 & 10.890  & 81 \\
		$3(s/d)_1$ & 11.084  & 100 & 11.086  & 95 \\  \hline
	\end{tabular}
	\caption{Spectrum of bottomonium ($S=0$) and bottomonium hybrids ($S=1$): $0^{-+}$ states. Masses are in GeV. The \% columns show the fraction of the hybrid components for the mass states in the previous column. The mixing potentials are fixed to $V_S^\Pi [+-]$ and $V_S^\Sigma [++]$.	$m_b=4.88$ GeV.}
	\label{b0-+}
\end{table}

\begin{table}[htbp]
	\centering
	\begin{tabular}{|c|c|c|c|c|}
		\hline
		$NL_J$ & $\lambda=0$ & \%   & $\lambda=0.6$  &\%    \\
		\hline 
		$1p$       & 9.908   &  0  & 9.886   &  2 \\ 
		$2p$       & 10.265  & 0   & 10.249  & 2  \\ 
		$3p$       & 10.553  & 0   & 10.543  & 0\\ 
		$1p_1$     & 10.761  & 100 & 10.761  & 100 \\ 
		$4p$       & 10.806  & 0   & 10.798  & 1 \\
		$1(p/f)_2$ & 10.819  & 100 & 10.820  & 100\\ 
		$2p_1$     & 10.970  & 100 & 10.969  & 100 \\
		$1p_0$     & 11.011  & 100 & 11.006  & 100 \\  \hline
	\end{tabular}
	\caption{Same as in table \ref{b0-+} for $1^{+-}$ states.}
	\label{b1+-}
\end{table}

\begin{table}[htbp]
	\centering
	\begin{tabular}{|c|c|c|c|c|}
		\hline
		$NL_J$ & $\lambda=0$ & \%   & $\lambda=0.6$  &\%    \\
		\hline 
		$1d$       & 10.155  &  0  & 10.144   &  2 \\ 
		$2d$       & 10.454  & 0   & 10.444  & 3  \\ 
		$1(s/d)_1$ & 10.690  & 100 & 10.685  & 82\\ 
		$3d$       & 10.712  & 0   & 10.717  & 52 \\ 
		$1d_2$     & 10.870  & 100 & 10.870  & 100 \\
		$2(s/d)_1$ & 10.885  & 100 & 10.886  & 94\\ 
		$1(d/g)_1$ & 10.935  & 100 & 10.937  & 99 \\
		$4d$       & 10.947  & 0   & 10.945  & 13 \\
		$2d_2$     & 11.074  & 100 & 11.074  & 99 \\  \hline
	\end{tabular}
	\caption{Same as in table \ref{b0-+} for $2^{-+}$ states.}
	\label{b2-+} 
\end{table}
	
	\end{appendices}

\end{document}